\documentclass[11pt]{article}
\usepackage{geometry}                
\usepackage{graphicx}
\usepackage{amssymb}
\usepackage{amsmath}
\usepackage{epstopdf}
\usepackage{amsthm}
\usepackage{threeparttable}
\usepackage{booktabs}
\usepackage{setspace}
\usepackage{xcolor}
\usepackage[ruled,vlined]{algorithm2e}
\newtheorem{theorem}{Theorem}

\newtheorem{assumption}{Assumption}

\newenvironment{remark}[1][Remark]{\begin{trivlist}
\item[\hskip \labelsep {\bfseries #1}]}{\end{trivlist}}

\def\lj38{{\rm LJ}_{38}}

 \title{Spectral analysis and clustering of large stochastic networks. \\Application to  the Lennard-Jones-75 cluster.}
 
 \author{Maria Cameron$^1$ and Tingyue Gan$^1$}

\begin{document}
\maketitle

 \footnotetext[1]{University of Maryland, 
 Department of Mathematics, 
 College Park, MD  20742,
 {\tt cameron@math.umd.edu, tgan@umd.edu}}
 
\abstract{
We consider stochastic networks with pairwise transition rates of the form $L_{ij} = a_{ij}\exp(-U_{ij}/T)$ where the temperature $T$ 
is a small parameter. Such networks arise in physics and chemistry
and serve as  mathematically tractable models of complex systems. Typically, such networks contain large number{s} of states and
widely varying pairwise transition rates.
We present a methodology for spectral analysis and clustering of such networks that takes advance of the small parameter $T$ and consists of two steps:   (1) computing zero-temperature asymptotics for 
eigenvalues and the collection of quasi-invariant sets, and (2) finite temperature continuation.
Step (1) is reducible to a sequence of optimization problems on graphs.
A novel single-sweep algorithm for solving them is introduced.
Its mathematical justification is provided.
This algorithm is valid for both time-reversible and time-irreversible networks.
For time-reversible networks, 
a finite temperature continuation technique combining lumping and truncation with Rayleigh quotient iteration 
is developed.  The proposed methodology is applied to the network representing the energy landscape of the Lennard-Jones-75 
cluster containing 169,523 states and 226,377 edges. The transition process between its two major funnels 
is analyzed. The corresponding eigenvalue  
is shown to have a kink at the solid-solid phase transition temperature.
}

\section{Introduction}
\label{sec:intro}
%
%
The contemporary development of communications, information technologies and powerful computing resources 
has made networks a popular tool for data organization, representation and interpretation. 
In particular, networks have demonstrated their strong potential for modeling the dynamics of complex physical systems. 
These include time-reversible processes such as 
atomic or molecular cluster rearrangements and conformal changes in molecules \cite{wales_book,wales_landscapes,web}, 
Markov State Models \cite{schuette_thesis,swope,pande05,pande07,noe07,noe09,prinz,schuette11},
as well as time-irreversible processes such as walks  of molecular motors \cite{astumian}.
Typically, such networks contain a large number of states or vertices ($n = 10^p$, $p=4,5,6,\ldots$), they are sparse and unstructured,
and the pairwise transition rates vary by tens of orders of magnitude. { As a result, analysis of the dynamics of such networks
is a challenging problem due to their size, complexity, and severe floating point arithmetic issues.} 

{ Several approaches quantifying  the dynamics of complex stochastic networks 
have been inroduced. A. Bovier and collaborators  developed the potential-theoretic approach and 
advanced the mathematical spectral theory of metastability \cite{bovier1,bovier2002,bovier_book}.
This theory is built upon an analogy with electric circuits. 
It is valid for time-reversible networks with an arbitrary form of pairwise transition rates.
Its important result is sharp estimates for small eigenvalues and corresponding eigenvectors obtained under the assumption of the existence
of spectral gaps. 
}

{ 
The Transition Path Theory (TPT)  originally proposed by W. E and E. Vanden-Eijnden \cite{eve1}
and further developed in \cite{dtpt,cve}, like the potential-theoretic approach, was inspired by an analogy with electric circuits.
However, there are important differences: $(i)$ TPT does not assume time-reversibility and 
$(ii)$ TPT is focused on the statistical analysis of so-called reactive trajectories.
An alternative approach for analyzing reactive trajectories 
based on a set of recurrence relationships was proposed by M. Manhart and  V. Morozov 
\cite{more1,more2}. 
}

{ 
Stochastic networks with pairwise transition rates of the form
\begin{equation}
\label{eq0}
L_{ij}=a_{ij}\exp(-U_{ij}/T),
\end{equation}
 where $a_{ij}>0$ and $U_{ij}>0$ are coefficients and $T$ is a small parameter (typically, the absolute temperature in the physical context)
 arise as coarse-grained models of continuous systems justified by the Large Deviation Theory \cite{f-w}.
M. Freidlin proposed to describe the dynamics of such networks on a set of long time scales by means of a 
hierarchy of cycles \cite{freidlin-cycles, freidlin-physicad,f-w} (we refer to them as Freidlin's cycles).
A. Wentzell developed asymptotic estimates for eigenvalues \cite{wentzell2}. 
In both cases, the results were given in the form of solutions of a series of
optimization problems on a certain kind of directed graphs called W-graphs. 
Time-reversibility was not assumed. Freidlin's and Wentzell's approaches were further advanced by E. Oliviery and M. Vares \cite{oval}.
}

{ 
The spectral analysis of stochastic networks with pairwise rates of the form of Eq. \eqref{eq0} 
was brought from the  theoretical field to the computational field in \cite{cspec1,cspec2} for the case of time-reversible networks.
The methodology presented in this work can be viewed as a two-fold extension of 
the one proposed in \cite{cspec1,cspec2}.
First, a novel single-sweep algorithm for finding asymptotic estimates for  
eigenvalues via solving the series of the optimization problems on W-graphs
is introduced. This algorithm does not require time-reversibility. 
Second,  the finite-temperature continuation technique based on the Rayleigh quotient iteration \cite{cspec2} is empowered 
by a series of truncations and lumpings. This makes it  
applicable to networks with an arbitrary range of pairwise transition rates.
}

The dynamics of a stochastic network (also known as a continuous-time Markov chain) 
is determined by its generator matrix $L$ and the initial probability distribution $p_0$. 
Throughout this paper, we assume that the spectral decomposition of $L$ exists, i.e.,
$L = \Phi Z\Phi^{-1}$, where $Z$ is a diagonal matrix with eigenvalues of $L$ along its diagonal, and columns of $\Phi$ are { the corresponding} 
right eigenvectors.
The spectral decomposition of the generator matrix 
gives a key to understanding the dynamics of the network, the extraction of quasi-invariant sets, 
and building coarse-grained models. 
However, its direct calculation for large and complex networks mentioned above might be exceedingly difficult due to issues 
related to floating-point arithmetic.  
Furthermore, even if the  spectral decomposition is successfully computed, its interpretation might be not straightforward.
Finally, for large networks, it is often desirable to extract only some particular eigenpairs 
associated with relaxation processes of interest rather than to obtain the whole set of eigenvalues and eigenvectors. 
Typically, the corresponding eigenvalues have relatively small, however, not necessarily the smallest, 
absolute values of their real parts. 
This means that, typically, the eigenvalues of physical interest are associated 
with some slowly but not necessarily the slowest decaying processes 
taking place in the system. 

These issues can be reconciled by using the following two-step approach. 
Step one is  the computation of  zero-temperature asymptotic estimates for eigenvalues and a collection of quasi-invariant sets.
The indicator functions of these quasi-invariant sets are asymptotic estimates for right eigenvectors in the time-reversible case \cite{bovier2002}.
We will show that each asymptotic eigenvalue is straightforward to interpret. 
Step two is the continuation of the eigenvalues  describing relaxation processes of physical interest to a range of finite temperatures.

\begin{figure}[htbp]
\begin{center}
\centerline{
(a)\includegraphics[width=0.2\textwidth]{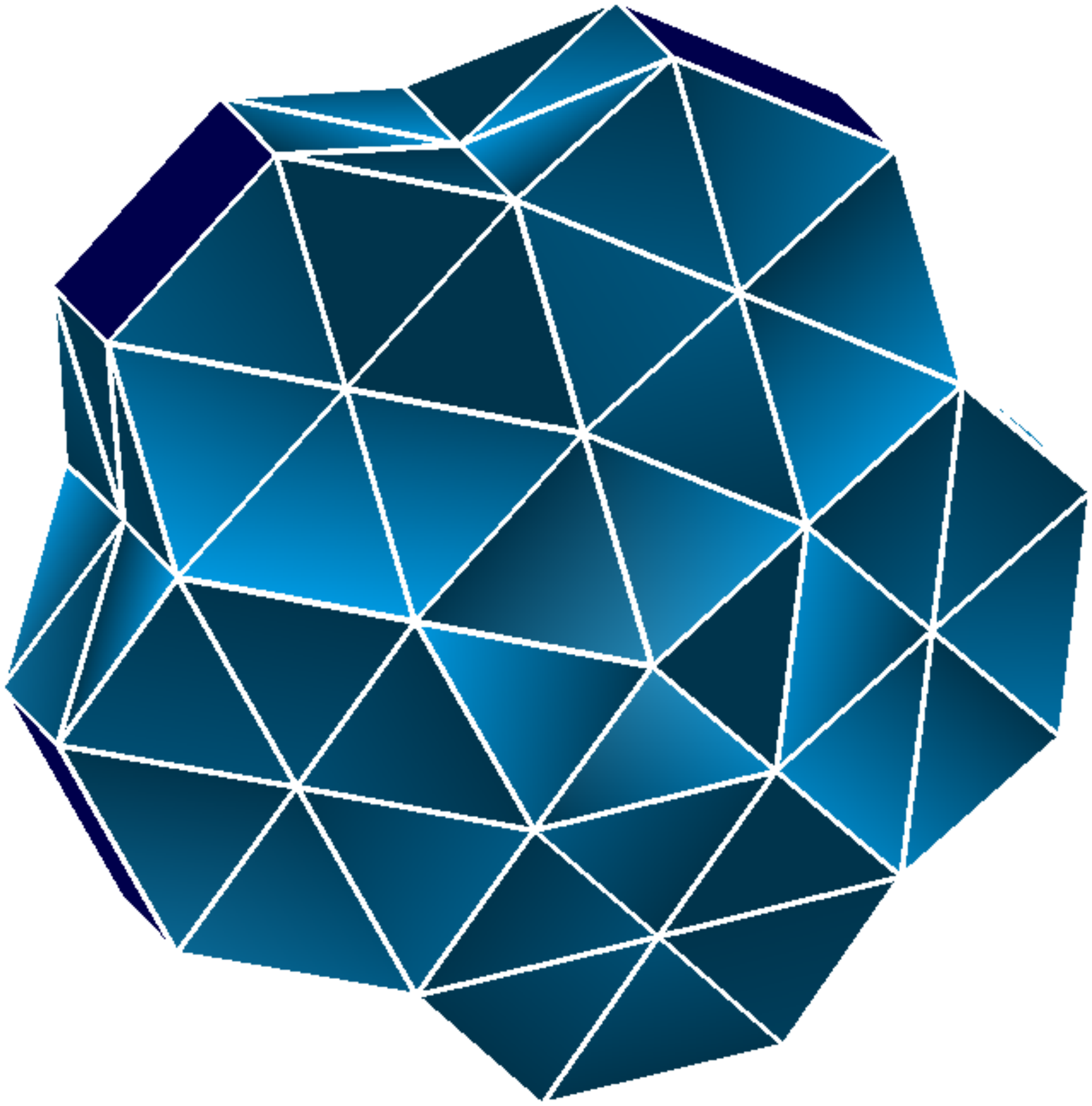}
(b)\includegraphics[width=0.2\textwidth]{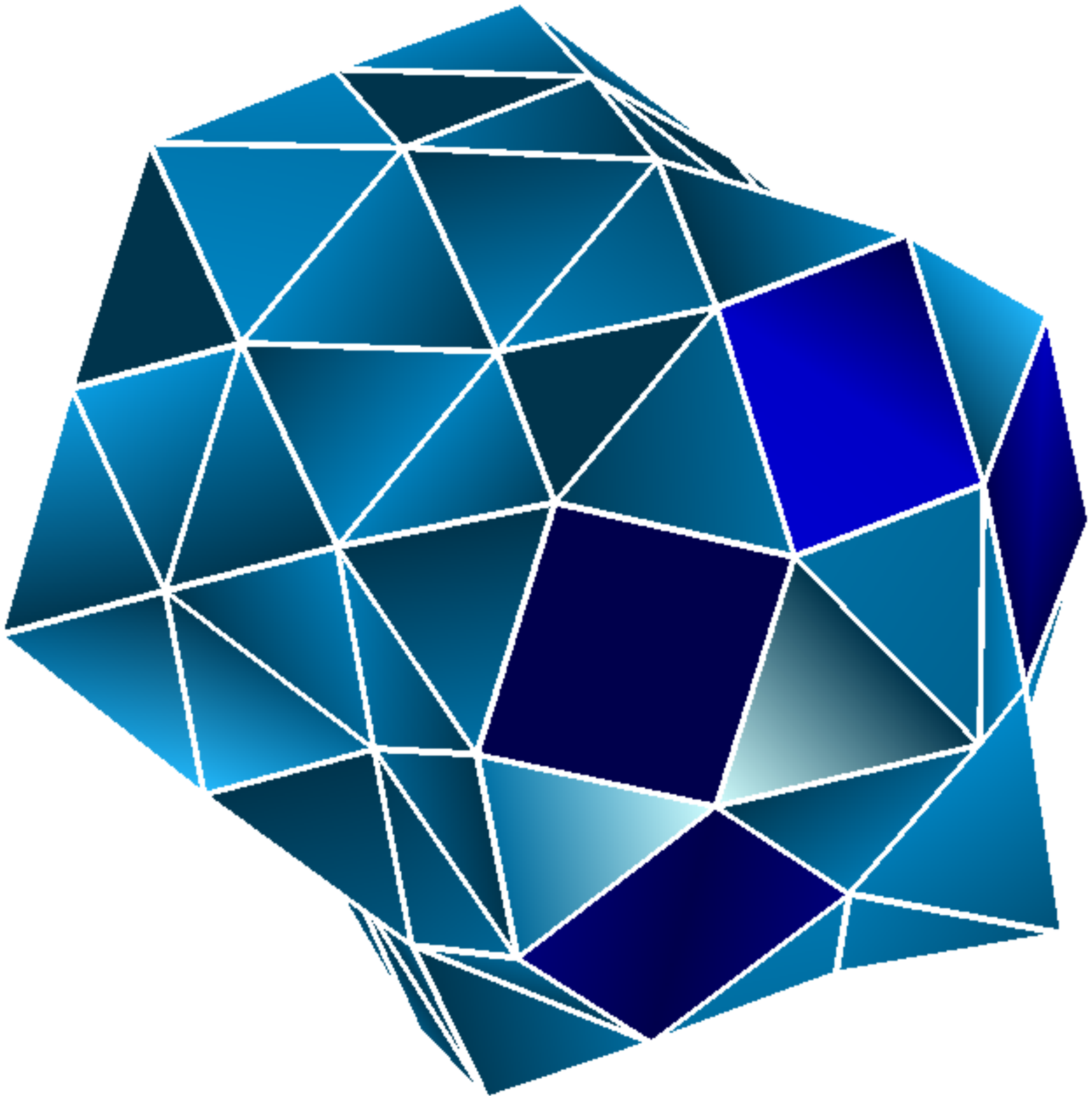}
}
\caption{(a): The global potential minimum (minimum 1) of the LJ$_{75}$ cluster, a Marks decahedron with the point group $D_{5h}$.
(b): The second lowest potential energy minimum (minimum 92), a configuration based on the icosahedral packing with the point group $C_1$ (no symmetry).}
\label{fig1}
\end{center}
\end{figure}

We apply the proposed approach to the stochastic network representing the energy landscape of the Lennard-Jones cluster of 75 atoms\footnotemark[1].
\footnotetext[1]{The data for the LJ$_{75}$ network were kindly provided by Professor D. Wales, Cambridge University, UK.}
For brevity, we will denote both, the Lennard-Jones cluster on $N$ atoms and the network representing its energy landscape, by LJ$_{N}$.
Vertices in this network correspond to local potential minima, 
while edges represent transition states between them. Transition rates between two adjacent minima
are given by the Arrhenius law.
{ Similar to} the well-studied LJ$_{38}$ cluster \cite{wales0,wales38,neirotti,picciani,cam1,cve,cspec1,cspec2}, 
the energy landscape of LJ$_{75}$ has a double-funnel structure (see  Fig. 5 in Ref. \cite{wales_landscapes} or Fig. 8.10(f) in Ref. \cite{wales_book}). 
The deep and narrow funnel is crowned with the global minimum which is the Marks decahedron with the point group $D_{5h}$ \cite{wales-doye}.
This is minimum 1 in Wales's data set.
{There are several local minima based on icosahedral packing at the bottom of the wide and shallower funnel}. 
One of them, minimum 92 in Wales's data set, is the second lowest one.
The two deepest minima are shown in Fig. \ref{fig1}.

Besides similarities, there are considerable differences between  LJ$_{38}$ and  LJ$_{75}$. 
First, the LJ$_{75}$ network is significantly larger than LJ$_{38}$ \cite{pathsample}. 
The LJ$_{75}$ dataset contains 593,320 local minima and 452,315 transition states. 
The largest connected component of this network, containing the two deepest potential minima, 1 and 92, consists of 
169,523 states (vertices) and 226377 undirected edges (excluding self-loops). 
The maximal vertex degree is 740. 
Second, LJ$_{75}$ 
has a wider range of potential barriers that presents
more severe numerical challenges than those encountered in the analysis of the LJ$_{38}$ network \cite{cve,cspec2}.
{ Third, unlike LJ$_{38}$,
LJ$_{75}$ has extremely low solid-solid transition critical temperature. 
This leads to an interesting phenomenon that we advertise below and discuss in details in Section \ref{sec:results}.}

Thermodynamic properties of  the LJ$_{75}$ cluster were studied in \cite{75solid-solid,mfc75}.
The process of physical interest in LJ$_{75}$ is the transition process between the two main funnels, 
the Marks decahedron funnel  and the icosahedral one.
The solid-solid transition critical temperature of LJ$_{75}$,
where the global potential energy 
minimum, minimum 1, gives place to icosahedral structures, 
is $T_{ss}^{{\rm LJ}75}=0.08$  reduced units \cite{75solid-solid}.
It  is very low  in comparison with 
the potential energy barrier of $7.897$ reduced units, separating minimum 1 from minimum 92. 
For comparison, the solid-solid transition critical temperature for LJ$_{38}$
is $T_{ss}^{{\rm LJ}38} = 0.12$  while the potential barrier separating the second lowest minimum  
from the global minimum is $3.543$ \cite{wales38,frantsuzov}. 
The melting temperature of the LJ$_{75}$, where another local maximum of the caloric curve is observed\footnotemark[2],  is $T_{sl}^{{\rm LJ}75}=0.25$.
 \footnotetext[2]{The value $T_{sl}^{{\rm LJ}75}=0.25$ corresponds to Wales's LJ$_{75}$ dataset containing 593,320 local minima.}
 Therefore, we will be interested in computing the eigenvalue and the eigenvector associated with the transition process between the two major 
 funnels for the temperature range $0\le T\le 0.25$. { For brevity, we will denote this eigenvalue by $\lambda(ICO-MARKS)$.}
 
{ The eigenvalue $\lambda(ICO-MARKS)$ approximately equals the transition rate between the two major channels of the energy landscape of LJ$_{75}$ \cite{bovier2002}.
 We will show that the graph of  $\log(\lambda(ICO-MARKS))$  versus $T^{-1}$ is nearly a piecewise-linear function
 with the kink at $T^{{\rm LJ}75}_{ss}=0.08$. 
 The slopes of the linear parts are in a good agreement with  
 theoretical predictions.
 This means that the solid-solid transition critical temperature
 $T^{{\rm LJ}75}_{ss}=0.08$ lies within the range where the Large Deviation regime is valid with a peculiar twist:
for $T>T^{{\rm LJ}75}_{ss}=0.08$, one needs to pretend that the icosahedral rather than the Marks decahedral funnel contains the global potential minimum.}
 
  The transition process between the two funnels is quantified by the corresponding eigencurrent.  
 We will show that, despite the 
 transition process becomes diverse as the temperature approaches the solid-liquid transition temperature $T_{sl}^{{\rm LJ}75}=0.25$,  
 one can extract a sequence of edges along which the eigencurrent is highly concentrated. 
 { Remarkably, this sequence is not a subsequence of the asymptotic
 zero-temperature path (the MinMax path) as it is in the case of the LJ$_{38}$ cluster \cite{cspec2}.}

 The rest of the paper is organized as follows. In Section \ref{sec:sign}, the significance of the spectral decomposition is discussed. 
In Section \ref{sec:evals}, the asymptotic estimates for eigenvalues are presented. In Section \ref{sec:nested}, nested properties of the optimal
W-graphs are formulated. In Section \ref{sec:evec}, asymptotic estimates for left and right eigenvectors
are discussed.  The single-sweep algorithm for computing zero-temperature asymptotic estimates for
eigenvalues and eigenvectors is introduced in Section \ref{sec:algorithm}. Section \ref{sec:LJ75}
is devoted to the application to LJ$_{75}$.  
The upgraded finite-temperature continuation technique is explained  in Section \ref{sec:LJ75finite}.
Concluding remarks and Acknowledgements are in Sections \ref{sec:conclusion} and \ref{sec:ac} respectively. 
The proof of the nested properties of the optimal W-graphs is found in the Appendix.

\section{Significance of the spectral decomposition}
\label{sec:sign}
\subsection{The general case}
Suppose we have a stochastic network (a continuous-time Markov chain) $(L,p_0)$, where $L$ 
is   the generator matrix, and $p_0$ is the initial probability distribution. 
The off-diagonal entries $L_{ij}$  of $L$ are the transition rates from states $i$ to states $j$, while the diagonal  entries
$L_{ii}$ are  defined so that the row sums of $L$ are zeros. The absolute values of the diagonal  entries $L_{ii}$ 
are the escape rates from states $i$.
Throughout this work we assume that the Markov chain has a finite number of states $n$ and is irreducible, 
i.e., there is a non-zero probability to reach any state from any other state. 
One can visualize the Markov chain using a directed 
weighted graph $G(S,A,L)$, where $S$ is its set of states (vertices), $|S|=n$, $A$ is its set of arcs (directed edges), 
and $L$ is the set of weights assigned to the arcs which is the set of all non-zero off-diagonal entries of the generator matrix $L$ (we abuse notations here).
Two vertices $i$ and $j$, $i\neq j$, are connected by an arc $(i\rightarrow j)$ if and only if $L_{ij}\neq 0$ (i.e., $L_{ij}>0$).

The time evolution of the probability distribution $p(t)=[p_1(t),\ldots,p_n(t)]$ is given by the master (or the Fokker-Planck) equation
\begin{equation}
\label{master}
\frac{dp}{dt} = pL,\quad p(0) = p^0.
\end{equation}
Its solution can be readily written in terms of the spectral decomposition of $L$, $L = \Phi Z \Psi$, where the columns of $\Phi$ are the right eigenvectors,
the rows of $\Psi{=\Phi^{-1}}$ are the left eigenvectors, and $Z$ is a diagonal matrix with the eigenvalues along its diagonal:
\begin{equation}
\label{eq1}
p(t)=p_0\Phi e^{tZ}\Psi = \sum_{k=0}^{n-1}(p_0\phi^k)e^{z_kt}\psi^k.
\end{equation}
In Eq. \eqref{eq1}, the solution $p(t)$ is expanded in the basis of the left eigenvectors $\psi_k$.
The coefficients of this expansion, $(p_0\phi^k)e^{z_kt}$, are the projections of the initial distribution $p_0$ onto the right eigenvectors $\phi^k$
multiplied by the exponential functions $e^{z_kt}$.
Given the above assumptions, it follows  from the Perron-Frobenius theorem that 
there is a unique eigenvalue $z_0\equiv\lambda_0=0$ of $L$, and the rest of the eigenvalues $z_k = -\lambda_k +i\mu_k$, $k=1,2,\ldots,n-1$,
have negative real parts. We order them so that
$$0=\lambda_0<\lambda_1\le\lambda_2\le\ldots\le\lambda_{n-1}.$$
The right eigenvector corresponding to $\lambda_0=0$ is $\mathbf{1} = [1,1,\ldots,1]^T$ as the row sums of $L$ are zeros, while left eigenvector
corresponding to $\lambda_0=0$ is the unique invariant probability distribution $\pi=[\pi_1,\ldots,\pi_n]$.
Using these notations and the fact that the distribution $p_0$ sums up to 1, we rewrite Eq. \eqref{eq1} as
\begin{equation}
\label{eq2}
p(t)=\pi +  \sum_{k=1}^{n-1}(p_0\phi^k)e^{-\lambda_kt}e^{i\mu_kt}\psi^k.
\end{equation}
Eq. \eqref{eq2} shows that  for any initial probability distribution $p_0$, the probability distribution $p(t)$ converges to the invariant distribution $\pi$,
 and 
 the decay rate for $k$-th eigencomponent of $p(t)$ is $\lambda_k$. Hence, on long time scales (say, $t>t^{\ast}$), the probability distribution
 $p(t)$ will be essentially determined  only by those eigencomponents where $\lambda_k$ is small (i.e., $\lambda_k < 1/t^{\ast}$).
 Since we are generally interested in the dynamics of the network on large times, it is important to be able to compute the
 eigenvalues with small real parts (in absolute values) and the corresponding left and right eigenvectors.

\subsection{Time-reversible Markov chains}
If the Markov chain is time-reversible, or, equivalently, 
its generator matrix $L$ is in detailed balance with the invariant distribution $\pi$, i.e., $\pi_iL_{ij} = \pi_jL_{ji}$, then 
the following additional properties of $L$ hold.
\begin{itemize}
\item
$L$ can be decomposed as 
\begin{equation}
\label{rev1}
L=P^{-1}Q,\quad {\rm where} \quad Q=Q^{T}~~ {\rm and}~~ P = \left[\begin{array}{ccc}\pi_1&&\\&\ddots&\\&&\pi_n\end{array}\right].
\end{equation}
\item
$L$ is similar to a symmetric matrix 
\begin{equation}
\label{rev2}
L_{sym} = P^{1/2}LP^{-1/2} = P^{-1/2}QP^{-1/2},
\end{equation}
therefore, its eigenvalues are real and non-positive.
\item
The matrices $\Psi$ and $\Phi$ are related via
\begin{equation}
\label{rev3}
\Psi = \Phi^TP,\quad \text{i.e.,}\quad \psi^k = (\phi^k)^TP = [\pi_1\phi_1^k,\ldots,\pi_n\phi_n^k],
\end{equation}
i.e., the right eigenvectors of $L$ are orthonormal with respect to the inner product weighted by the invariant distribution, i.e., $\Phi^TP\Phi = I$.
\end{itemize}

Relaxation processes in time-reversible Markov chains can be quantitatively described in terms of eigencurrents.
A detailed discussion on it is found in \cite{cspec2}; also see \cite{kurchan6,eve2014,cve}.
Taking into account Eq. \eqref{rev3}, the time evolution of the $i$-th component of the probability distribution can be rewritten as
\begin{equation}
\label{rev4}
\frac{dp_i}{dt} = (pL)_i = \sum_{j=1}^np_jL_{ji} = \sum_{j\neq i}(p_jL_{ji}-p_iL_{ij}).
 \end{equation}
Plugging  Eq. \eqref{rev3} in and using the detailed balance condition $\pi_iL_{ij}=\pi_jL_{ji}$ we get
 \begin{equation}
 \label{rev5}
\frac{dp_i}{dt} =  - \sum_{k=0}^{n-1}(p_0\phi^k)e^{-\lambda_kt}\sum_{j\neq i}\pi_iL_{ij}(\phi^k_i - \phi_j^k) =  - \sum_{k=0}^{n-1}(p_0\phi^k)\sum_{j\neq i} F^k_{ij},
\end{equation}
where
\begin{equation}
\label{rev6}
F^k_{ij}(t): = e^{-\lambda_kt}\pi_iL_{ij}(\phi^k_i - \phi_j^k) 
\end{equation}
is the $k$-th eigencurrent along the arc $(i\rightarrow j)$.
Note that  $c_kF^k_{ij}$ is the expectation of the difference {of} the
numbers of transitions from $i$ to $j$ and from $j$ to $i$ {performed} by the system per unit time at time $t$ if the initial distribution is 
$$
\pi + c_k\psi^k = \left(\pi_1(1+c_k\phi^k_{1}),\ldots,\pi_n(1+c_k\phi^k_n)\right).
$$
({The coefficient $c_k$ is introduced} to ensure that all entries of $\pi + c_k\psi^k$ are nonnegative.)

Now we go over some properties of eigencurrents. 
Since $\phi^0_i= 1$, $i\in S$,  { the eigencurrent $F^{0}$ associated with the zero eigenvalue $z_0=0$ is zero everywhere. 
Furthermore, $F^k_{ij} = - F^{k}_{ji}$, as immediately follows from the definition}.
It is easy to verify that the sum of eigencurrent $F^k$ at vertex $i$ is
\begin{equation}
\label{rev7}
\sum_{j\neq i} F^k_{ij} = e^{-\lambda_kt}\lambda_k\pi_i\phi^k_i.
\end{equation}
Therefore, the eigencurrent $F^k$ is not, in general, conserved at vertex $i$ but is either emitted, if $\phi^k_i>0$, or absorbed, if $\phi_i^k < 0$.
Hence, the set of states can be partitioned into emitting (more precisely, non-absorbing) and absorbing states: $S = S_{+}^k\cup S_{-}^k$, where
$$
S^k_{+}: = \{i\in S~:~\phi^k_i\ge 0\}\quad{\rm and}\quad S_{-}^k = \{i\in S~:~\phi^k_i <  0\}.
$$
This partition of the network  is called  \emph{the $k$-th emission-absorption cut} or, briefly,  \emph{the $k$-th EA-cut}. The set of 
edges with endpoints in different components of this partition is also called the EA-cut.

It is shown in \cite{cspec2} that out of all cuts of the given network, the total eigencurrent  $F^k$ flowing through the edges of the $k$-th EA-cut
is maximal.  It is convenient to normalize the eigencurrent $F^k$ so that its flux through the $k$-th EA-cut is unit or 100\%. 
Then one can find the percentages of the eigencurrent $F^k_{ij}$ for every edge $(i,j)$ and hence obtain a quantitative description 
of the relaxation process starting from the initial distribution $\pi + c_k\psi^k$. 
{ In Section \ref{sec:ecur}, we will analyze} the eigencurrent  corresponding to the transition process between the
two major funnels
of LJ$_{75}$.

\section{Asymptotic estimates for eigenvalues}
\label{sec:evals}
\begin{figure}[htbp]
\begin{center}
\includegraphics[width=0.6\textwidth]{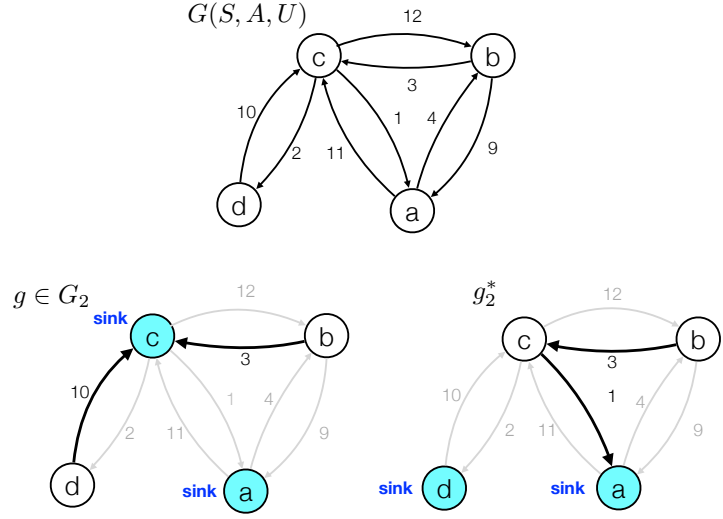}
\caption{An example of a graph $G(S,A,U)$ (top) and two W-graphs for $G(S,A,U)$ with two sinks (bottom):  a W-graph with two sinks (bottom left)
and the optimal W-graph with two sinks (bottom right).}
\label{fig2}
\end{center}
\end{figure}

A relationship between the zero-temperature asymptotics for eigenvalues of the generator matrix $L$ 
with off-diagonal entries of the order of $L_{ij}\asymp \exp(-U_{ij}/T)$ and { a series of optimization problems}
on W-graphs was established by A. Wentzell in 1972  \cite{wentzell2,f-w}.
A W-graph with $k$ sinks for the given directed weighted graph $G(S,A,U)$ is defined as a subgraph of $G$
obtained as follows:
(a) select a set of $k$ vertices and call them sinks; (b) choose a subset of $n-k$ arcs (directed edges) 
so that $(i)$ there is a single outgoing arc from every non-sink vertex and $(ii)$ the graph has no cycles. 
Note that a W-graph with $k$ sinks contains $n-k$ arcs.  In particular, the W-graph with $n$ sinks has no arcs.
An optimal W-graph with $k$ sinks, denoted by $g^{\ast}_k$, is the one for which the sum 
of weights of its arcs is minimal possible.  I.e.,
\begin{equation}
\label{owg}
g^{\ast}_k = \arg\min_{g\in G_k}\sum_{(i\rightarrow j)\in g}U_{ij},\quad 1\le k\le n,
\end{equation}
where $G_k$ is the set of all W-graphs on $G(S,A,U)$ with $k$ sinks.
An example of a graph with four vertices and some of its W-graphs with two sinks are shown in Fig. \ref{fig2}.

Wentzell's asymptotic estimates for eigenvalues  are up to the exponential order  \cite{wentzell2,f-w}.
For networks with pairwise transition rates of the form of Eq. \eqref{eq0}, i.e., $L_{ij} = a_{ij}\exp(-U_{ij}/T)$,
Wentzell's result can be upgraded so that the estimates for the eigenvalues include not only exponents but also pre-exponential factors (pre-factors).
Let us define the graph $G(S,A,U)$ for the given  generator matrix $L$, where the set of weights $U=\{U_{ij}\}$ is the set of the exponential coefficients 
from Eq. \eqref{eq0}. 
{Throughout the rest of the paper we will adopt the following \emph{genericness assumption}:
\begin{assumption}
All optimal W-graphs $g^{\ast}_k$, $1\le k\le n$, for the graph $G(S,A,U)$ are unique.
\end{assumption} }
Then all eigenvalues of $L$ are real and distinct and are approximates by
\begin{align}
\lambda_k &= A_ke^{-\Delta_k/\epsilon}(1+o(1)),\quad {\rm where} \label{eq3} \\ 
\Delta_k &= \sum_{(i\rightarrow j)\in g^{\ast}_k} U_{ij} - \sum_{(i\rightarrow j)\in g^{\ast}_{k+1}} U_{ij},\label{eq4}\\ 
A_k& = \frac{\prod_{(i\rightarrow j)\in g^{\ast}_k}a_{ij}}{\prod_{(i\rightarrow j)\in g^{\ast}_{k+1}}a_{ij}}. \label{eq5}
\end{align}
Below we provide a sketch of the proof of Eqs. \eqref{eq3}-\eqref{eq5}.

Eqs. \eqref{eq3}-\eqref{eq5} are obtained from the consideration of the characteristic polynomial
$$
\rho_L(z) = z^n + \beta_{n-1}z^{n-1} +\beta_{n-2}z^{n-2} + \ldots + \beta_1z = z(z-z_1)\ldots(z-z_{n-1})
$$ 
of the generator matrix $L$. 
It follows from our construction of the optimal W-graphs by Algorithm 1 presented in Section \ref{sec:pse} below, that 
the uniqueness assumption of the optimal W-graphs implies that 
$$
0<\Delta_{n-1}<\Delta_{n-2}<\ldots<\Delta_1.
$$
For this case, it is  proven in \cite{wentzell2} that all eigenvalues are real and distinct, i.e., 
$z_k = -\lambda_k $, where $\lambda_k >0$, $1\le k\le n-1$.
One can show that the coefficients {of the characteristic polynomial}  $\rho_L(z)$ are given by
\begin{equation}
\label{beta}
\beta_{n - k} = \sum_{1\le i_1\neq\ldots\neq i_{k}\le n-1} \prod_{l=1}^{k}\lambda_{i_l} =  
\sum_{g\in G_{k}}\left(\prod_{(i\rightarrow j)\in g}a_{ij}e^{-U_{ij}/T}\right).
\end{equation}
The summands in Eq. \eqref{beta} range exponentially. Hence, each sum is dominated by its largest term. 
In the sum over the  W-graphs with
$k$ sinks, the largest term is achieved on the optimal W-graph $g^{\ast}_k$ which is unique by our assumption. 
The sum of products of all combinations of $k$  numbers
$\lambda_{i_l}$, $1\le l\le k$,  where all $1\le i_1,\ldots, i_k\le n-1$  are distinct, is dominated by $\lambda_{n-1}\ldots \lambda_{n-k}$.
Therefore, 
$$
\lambda_k = \frac{\lambda_{n-1}\ldots \lambda_{k}}{\lambda_{n-1}\ldots \lambda_{k+1}} \approx
\frac{ \prod_{(i\rightarrow j)\in g^{\ast}_k} a_{ij}e^{-U_{ij}/T}}{\prod_{(i\rightarrow j)\in g^{\ast}_{k+1} }a_{ij}e^{-U_{ij}/T}},
$$
{ and  hence Eqs. \eqref{eq3}-\eqref{eq5} hold.}

For the time-reversible networks, the asymptotic estimates for the eigenvalues can be simplified.
In this case,  we assume that the off-diagonal entries of $L$ are of the form 
\begin{equation}
\label{eq6}
L_{ij} = \frac{b_{ij}}{b_i}e^{-(V_{ij} - V_i)/T},\quad{\rm where}\quad V_{ij} = V_{ji}~~{\rm and}~~b_{ij} = b_{ji}~~\text{for all}~~i,j\in S.
\end{equation}
In other words, there exist the potential $V$ and the pre-factor function $b$
defined on all vertices and all edges such that $U_{ij} = V_{ij} - V_i$ and $a_{ij} = b_{ij}/b_i$.
Then the invariant probability distribution is readily calculated and given by
\begin{equation}
\label{eq7}
\pi_i = b_ie^{-V_i/T},\quad 1\le k\le n.
\end{equation}
It was proven in \cite{cspec1} that for stochastic networks with pairwise rates of the form of Eq. \eqref{eq6}, 
all edges (with erased directions) belonging to the optimal W-graph $g^{\ast}_{k+1}$, also belong to
the optimal  W-graph $g^{\ast}_k${,} and all sinks of $g^{\ast}_k$ are also sinks of $g^{\ast}_{k+1}$. Hence, $g^{\ast}_k$ is obtained from $g^{\ast}_{k+1}$
by removing one sink and adding one edge. We denote the disappearing sink and the newly 
added edge by $s^{\ast}_k$ and $(p_k^{\ast}\rightarrow q^{\ast}_k)$ respectively.
Taking all this into account, it is easy to calculate that the exponent and the pre-factor for the 
asymptotic estimate of the $k$-th eigenvalue $\lambda_k$:
\begin{equation}
\label{eq8}
\Delta_k = V_{p^{\ast}_kq^{\ast}_k}-V_{s^{\ast}_k},\quad A_k = \frac{b_{p^{\ast}_kq^{\ast}_k}}{b_{s^{\ast}_k}}.
\end{equation}

\section{Nested properties of optimal W-graphs}
\label{sec:nested}
\begin{figure}[htbp]
\begin{center}
\includegraphics[width = 0.6\textwidth]{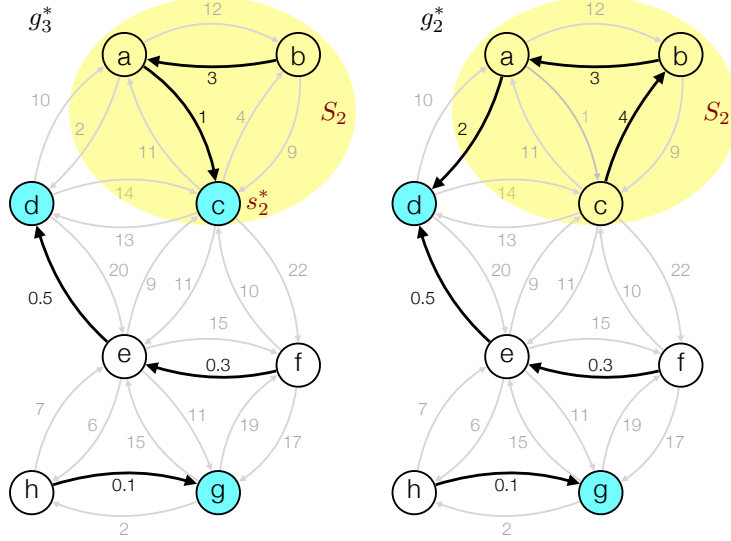}
\caption{An illustration for the nested properties (Theorem \ref{the:nested}). }
\label{fig:nested}
\end{center}
\end{figure}

As we have discussed in Section \ref{sec:evals}, in order to obtain asymptotic estimates 
for eigenvalues {of} the generator matrix $L$ of size $n\times n$,
one needs to find the collection of optimal W-graphs $g^{\ast}_k$, 
$1\le k\le n$, i.e., to solve the collection of optimization problems on graphs given by Eq. \eqref{owg}.
In order to find an optimal W-graph $g^{\ast}_k$ with $k$ sinks, one needs to minimize the sum in Eq. \eqref{owg}
with respect to $(i)$ the choice of $k$ sinks out of $n$ vertices,  and $(ii)$ the choice of arcs so that each non-sink vertex
has exactly one outgoing arc and these arcs create no cycles. 
Optimization problem \eqref{owg} is hard to solve by brute force because
the number of W-graphs for large number of vertices $n$ is typically enormous.
Fortunately, the collection of optimal $W$-graphs $\{g^{\ast}_k\}_{k=1}^n$ 
under the assumption that all
of them are unique, possesses nested properties allowing us to find them recursively starting from $g^{\ast}_n$ and 
finishing with $g^{\ast}_1$.
\begin{theorem} \emph{(Nested properties of optimal W-graphs)}
\label{the:nested}
Let $G(S,A,U)$ be a weighted directed graph  with $n$ vertices such  that all optimal W-graphs $g^{\ast}_k$, $1\le k\le n$
are unique. Then  for all $1\le k<n$ the following nested properties hold.
\begin{enumerate}
\item
Every sink of $g^{\ast}_k$ is also a sink of $g^{\ast}_{k+1}$;
\item
Let $S_k$ be the set of vertices in the connected component  of  $g^{\ast}_{k+1}$ containing no sink of $g^{\ast}_{k}$.
The collection of  outgoing arcs  from the subset of vertices $S\backslash S_k$ in $g^{\ast}_{k+1}$ 
 coincides with  the collection of outgoing arcs from the subset of vertices $S\backslash S_k$
in $g^{\ast}_k$. 
\item
There is a single outgoing arc in $g^{\ast}_k$ with tail in $S_k$ and head in $S\backslash S_k$.
\end{enumerate}
\end{theorem}
\begin{remark}
The collections of outgoing arcs from the subset of states $S_k$  in $g^{\ast}_{k+1}$ 
and $g^{\ast}_k$ do not necessarily coincide.
\end{remark}
A proof of Theorem \ref{the:nested} is found in the Appendix. The nested properties stated in Theorem \ref{the:nested} are illustrated in Fig. \ref{fig:nested}. 
The sets of arcs of the optimal W-graphs $g^{\ast}_3$ (left)  and $g^{\ast}_2$ (right) with tails not in $S_2=\{a,b,c\}$ coincide. All sinks of $g^{\ast}_2$
 are also sinks of $g^{\ast}_3$. There is a single outgoing arc $(a\rightarrow d)$ in $g^{\ast}_2$ with tail in $S_2$ and head in $S\backslash S_2 = \{d,e,f,g,h\}$.

Theorem \ref{the:nested} can be compared to Theorem 3.5 in \cite{cspec1} stating 
the nested properties of the optimal W-graphs for the time-reversible
case. In the time-reversible case, where $U_{ij} = V_{ij} - V_i$, $V_{ij}=V_{ji}$, $i,j\in S$, 
Claims 2 and 3 of Theorem \ref{the:nested} can be amplified as follows. 
Let $\mathcal{T}_{k}^{\ast}$ and $\mathcal{T}_{k+1}^{\ast}$ be the optimal forests obtained from the optimal W-graphs $g^{\ast}_k$ and $g^{\ast}_{k+1}$
by erasing the directions of their arcs. Then the forest $\mathcal{T}_{k+1}^{\ast}$ is a sub-forest of $\mathcal{T}_{k}^{\ast}$.
In addition, all optimal forests $\mathcal{T}_{k}^{\ast}$, $1\le k\le n${,} are subgraphs of the minimum spanning tree 
$$
\mathcal{T}^{\ast} = \arg\min_{\mathcal{T}}\sum_{(i,j)\in \mathcal{T}}V_{ij},
$$
(see Theorem 3.4 in \cite{cspec1}).
Here, the minimum is taken over all spanning trees for the graph $G(S,E,V)$, where vertices $i$ and $j$ are connected by an edge $(i,j)$ whenever
there is an arc $(i\rightarrow j)$ or $(j\rightarrow i)$ in $G(S,A,U)$, and  the weights 
$V_{ij}$ are taken from $U_{ij} = V_{ij}-V_i$.

\section{Asymptotic estimates for eigenvectors}
\label{sec:evec}
In this paper, we limit our discussion on asymptotic estimates for right and left eigenvectors to the time-reversible case.
Sharp asymptotic estimates for the right eigenvectors for time-reversible Markov chains were 
obtained in \cite{bovier2002} (see also \cite{bovier1}). 
It follows from the theory developed in \cite{bovier2002}  that 
the $k$-th right eigenvector of $L$ is approximated by the indicator vector $\phi_0^k =   [\phi^k_0(1),\ldots,\phi^k_0(n)]^T$
 of the quasi-invariant set $S_k$ defined in Theorem \ref{the:nested}, i.e.
\begin{equation}
\label{evec}
 \phi_0^0=\left[\begin{array}{c}1\\\vdots\\1\end{array}\right],\quad \phi_0^k(i) = \begin{cases}1,&i\in S_k\\0,&i\notin S_k\end{cases},~~1\le k\le n-1.
\end{equation}


Asymptotic estimates for left eigenvectors of $L$ can be deduced from those for the right ones.
Since the matrix whose rows are left eigenvectors is the inverse of the matrix whose columns 
are the corresponding right eigenvectors, we expect that
the matrix of the asymptotic estimates for left eigenvectors is the inverse of the matrix 
$$
\Phi_0:=[\phi_0^0~\ldots ~\phi_0^{n-1}].
$$
Below we will define a matrix $\Psi_0$ and show that it is the inverse of $\Phi_0$.

Let $S_k$ be the set of vertices in the connected component of the optimal W-graph $g^{\ast}_{k+1}$
containing no sink of $g^{\ast}_k$. 
Let $s^{\ast}_k$ be the sink of $S_k$. 
According to Theorem \ref{the:nested}, there exists a unique arc $(p^{\ast}_k,q^{\ast}_k)$ in $g^{\ast}_k$
such that $p^{\ast}_k\in S_k$ and  $q^{\ast}_k\in (S\backslash S_k)$. Let $T_k$ be the connected component of $g^{\ast}_{k+1}$
such that $q^{\ast}_k\in T_k$. Its sink will be denoted by $t^{\ast}_k$ respectively.
In the example in Fig. \ref{fig:nested}, $k=2$, $S_2 = \{a,b,c\}$, $s^{\ast}_2=c$, $(p^{\ast}_2,q^{\ast}_2) = (a,d)$, 
$T_2 = \{d,e,f\}$, and $t^{\ast}_2 = d$.

We introduce  
row vectors  $\psi^k_0=[\psi^k_0(1),\ldots,\psi^k_0(n)]$ { given by}
\begin{equation}
\label{evec1}
\psi_0^k(i) = \begin{cases}1,&i=s^{\ast}_k\\-1,&i=t^{\ast}_k\\0,&\text{otherwise},\end{cases}, ~~ 1\le k\le n-1,\quad 
\psi_0^0(i) = \begin{cases}1,&i=s^{\ast}_0\\0,&\text{otherwise}.\end{cases}
\end{equation}

Let $\Psi_0$ be the matrix with rows $\psi_0^k$. We claim that $\Psi_0$ is the inverse of $\Phi_0$, i.e.,
\begin{equation}
\label{or}
\Psi_0\Phi_0 = \Phi_0\Psi_0 = I.
\end{equation}
Let us prove Eq. \eqref{or}. We have
\begin{equation}
\label{p1}
\psi_0^0\phi_0^r  =  \phi^r_0(s^{\ast}_0)  = \begin{cases} 1,& r = 0,\\0,& 1\le r\le n-1,\end{cases}
\end{equation}
and
\begin{equation}
\label{p2}
\psi_0^k\phi_0^r  =  \phi^r_0(s^{\ast}_k)  - \phi^r_0(t^{\ast}_k),\quad 1\le k \le n-1, ~0\le r\le n-1,\\
\end{equation}
If $ k = r$,  $s^{\ast}_k\in S_k$ and $t^{\ast}_k\notin S_k$. Hence 
$$
\psi_0^k\phi_0^k = 1,\quad 1\le k\le n-1.
$$
If $k<r$ then $s^{\ast}_k\notin S_r$ and $t^{\ast}_k\notin S_r$. Hence for all $i\in S$, $\phi^r_0(s^{\ast}_k) =  \phi^r_0(t^{\ast}_k)= 0$, 
If $k>r$ then the sinks $s^{\ast}_k$ and $t^{\ast}_k$ lie in the same connected component of $g^{\ast}_{r+1}$. 
Hence  $\phi^r_0(s^{\ast}_k) = \phi^r_0(t^{\ast}_k)$. Therefore, for $k\neq r$, Eq. \eqref{p2} implies that $\psi_0^k\phi_0^r = 0$.
This completes the proof.

 \section{The single-sweep algorithm}
 \label{sec:algorithm}
 In this Section, we introduce  a single-sweep algorithm for computing the collection of optimal W-graphs,
 suitable for both time-reversible and time-irreversible networks. The presentation of this algorithm is
 given in the form convenient for programming.

 \subsection{Building the hierarchy of the optimal W-graphs}
 We will build the hierarchy of the optimal W-graphs $g^{\ast}_k$ from bottom to top, i.e., starting from $g^{\ast}_{n}$ containing 
 $n$ sinks (every vertex is a sink of $g^{\ast}_n$) and no arcs, and finishing with $g^{\ast}_1$ containing $n-1$ arcs and one sink.
 Due to the nested properties (see Theorem \ref{the:nested}) we know that we will need to remove one sink in a time, possibly rearrange arcs
in the connected component containing the sink being removed, and connect it to some other connected component with a single arc.
The algorithm with all its nuances will be written in the form of a pseudocode in Section \ref{sec:pse}. 
In this Section,
we will explain some crucial parts of the algorithm such as the next arc selection procedure 
and the update rule \eqref{update_rule} for weights $U_{ij}$  and pre-factors $a_{ij}$.
 
In a  W-graph, there can be at most one outgoing arc from each vertex. 
Therefore, in order to find the hierarchy of the optimal W-graphs, we start with picking the
outgoing arcs of minimum weight from each vertex.  We proceed as follows.
Let $B_i$ be the set of outgoing arcs from the vertex $i$, $1\le i\le n$. 
In each set $B_i$, sort the arcs according to their weights $U_{ij}$,
find the outgoing arc of minimal weight, and denote it by \verb|min_arc|$(i)$. Then define the set 
$$
M := \bigcup_{i\in S}\verb|min_arc|(i).
$$
Now we start building optimal W-graphs.
Find the arc $(i\rightarrow j)$ of minimal weight in the set $M$, remove it from $M$,  and define the set $G^{\ast} = \{(i\rightarrow j)\}$ of
arcs  that has been removed from $M$. 
The arcs from $G^{\ast}$ will be used for the construction of optimal W-graphs.
Define the optimal W-graph $g^{\ast}_{n-1}$ with one arc $(i\rightarrow j)$
and $n-1$ sinks  $S\backslash\{i\}$.  
Then find the next minimal weight arc $(x\rightarrow y)$ in $M$, remove it from $M$ and add to $G^{\ast}$.
Suppose  $(x\rightarrow y)\neq (j\rightarrow i)$, i.e., 
it does not create a cycle with the arc $(i\rightarrow j)$.
Then define the optimal W-graph $g^{\ast}_{n-2}$ 
with two arcs $(i\rightarrow j)$ and $(x\rightarrow y)$
and $n-2$ sinks $S\backslash\{i,x\}$.
Suppose that, proceeding in this {manner}, the minimal weight arcs removed from $M$ are such that they do not form any cycles.
In this case, we will build the whole hierarchy of optimal W-graphs in $n-1$ steps, and the numbers $\Delta_k$ and $A_k$
defining the asymptotic estimates for the eigenvalues will be 
$$
\Delta_k = U_{p_{k}^{\ast},q_{k}^{\ast}},\quad A_k = a_{p_{k}^{\ast}q_{k}^{\ast}},\quad n-1\ge k\ge 1.
$$

However, such a scenario can occur only if we are particularly lucky with the network. Typically, sooner or later, the next minimal weight arc
removed from the set $M$ will create a directed cycle with some arcs in $G^{\ast}$. Hence, we cannot define the next optimal W-graph
by adding this arc to the previous one. 
We need to develop another procedure to handle this case. 

\begin{figure}[htbp]
\begin{center}
\includegraphics[width=0.6\textwidth]{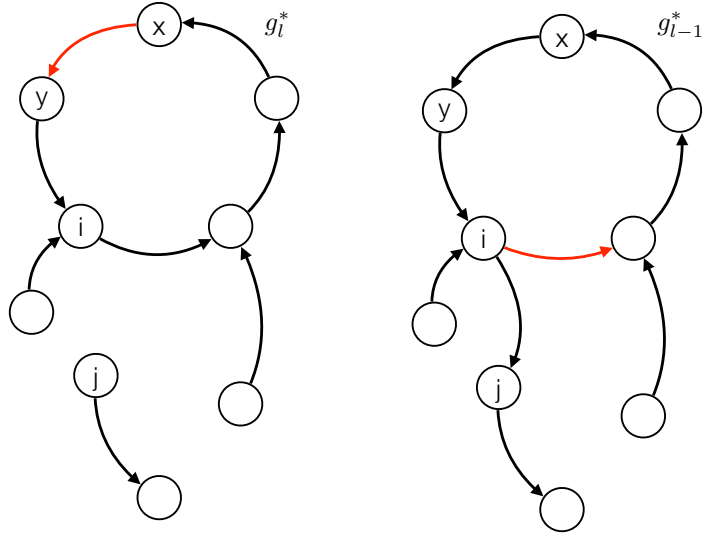}
\caption{An illustration for the arc replacement in a cycle and the update rule \eqref{update_rule}. 
Black and red arcs belong to the set $G^{\ast}$. Black arcs form 
optimal W-graphs. The arc $(x\rightarrow y)$ creates a cycle with other arcs in $G^{\ast}$.}
\label{fig:cycle}
\end{center}
\end{figure}

Suppose { the
optimal W-graphs $g^{\ast}_{n}$, ..., $g^{\ast}_{k}$  are constructed}
as described above.
Let the endpoints of the next minimal weight arc $(x\rightarrow y)$ removed from $M$ 
lie in the same connected component of $g^{\ast}_k$. Since, by construction, the set $M$ contains at most one outgoing arc
from each vertex, the new arc $(x\rightarrow y)$ 
creates a simple directed cycle $C$ with some other arcs in $g^{\ast}_k$:
$$
C=\{x\rightarrow y\rightarrow \ldots\rightarrow x\}.
$$
{Observe that, after adding arc $(x\rightarrow y)$, every vertex in the connected component $S_C$ of $g^{\ast}_k$
containing the cycle $C$ has an outgoing arc. Furthermore, observe that $x$ is a sink of $g^{\ast}_k$, as,  prior to adding the arc $(x\rightarrow y)$,
the vertex $x$ have had no outgoing arc in $g^{\ast}_k$.
In order to merge the component $S_C$  }
with some other connected component,
we need to  add  an arc $(i\rightarrow j)$  such that  $i\in C$ and  $j\in(S\backslash C)$ to the set $M$. 
{The fact that  $i$ must lie in $C$ follows from Theorem \ref{the:nested}.}
Indeed, suppose that the arc $(i\rightarrow j)$ is  removed from $M$  
after the optimal W-graph $g^{\ast}_l$ for some $l < k$ has been built, while $g^{\ast}_{l-1}$ has not been built yet.
First, for simplicity, we assume that no other cycle except for $C$ has been created up to this point.
Then, in order to build $g^{\ast}_{l-1}$, we will replace the arc $ \verb|min_arc|(i)\in C$ with the arc $(x\rightarrow y)\in C$ and add the arc $(i\rightarrow j)$
as shown in Fig. \ref{fig:cycle}.
Then the connected component of $g^{\ast}_l$ containing $i$ will remain connected in $g^{\ast}_{l-1}$, 
and every vertex
will have at most one outgoing arc in $g^{\ast}_l$.
The addition of the arc $(i\rightarrow j)$ increases the sum of  weights by $U_{ij} + U_{xy}-U_{\verb|min_arc|(i)}$.

Motivated by these considerations, we update the weights  and pre-factors of the outgoing arcs with tails in the cycle $C$ according to the
following update rule: for all $(i\rightarrow j)\in B_i$, for all $i\in C$ except for $x$, we set
\begin{equation}
\label{update_rule}
U_{ij} = U_{ij} + U_{xy}-U_{\verb|min_arc|(i)} ,\quad
a_{ij} =  a_{ij}  a_{xy}/a_{\verb|min_arc|(i)} .
\end{equation}
Note that we do not need to update the weights of the arcs emanating from $x$.
Indeed, the last added arc with tail at $x$ is $x\rightarrow y\equiv \verb|min_arc|(x)$.  
Hence the update rule \eqref{update_rule} will not change
the weights and pre-factors of the arcs emanating from $x$.
Then we merge all sets $B_i$ where $i\in C$ into one set $B$.
Next, we select the arc $(p\rightarrow q)\in B$ of minimum weight (the effective weight),  remove it from $B$, add it to $M$, and
for each $i\in C$ we set \verb|min_arc|$(i) = (p\rightarrow q)$ and $B_i=B$.
After that, we again remove the arc of minimum weight from $M$ and add it to $G^{\ast}$. 
 
Suppose { that again}, the next minimal weight arc $(v\rightarrow w)$ removed from $M$ creates a simple directed cycle $C$
with some other arcs  in $G^{\ast}$. 
 We update the arc weights in all sets $B_i$, $i\in C$, according to the update rule \eqref{update_rule}
 and merge them to one set $B$. 
We call all previously created cycles containing vertices of $C$ sub-cycles of $C$.
Recursively,  all sub-cycles of any sub-cycle  of $C$ will be also called sub-cycles of $C$. 
We denote the union of the cycle $C$ and all its sub-cycles by $C'$.
Note that, by induction, 
the set $B$ is the union of all arcs emanating from all vertices of the cycle $C'$. 
Then we find the arc $(p\rightarrow q)$ of minimal weight in $B$, remove it from $B$, add it to $M$, and set 
$\verb|min_arc|(i) = (p\rightarrow q)$ and $B_i=B$ for all $i\in C$.

Proceeding according to these recipes, we build the whole hierarchy of the optimal W-graphs $g^{\ast}_k$ and 
obtain the collection{s} of numbers $\Delta_k$, $A_k$, collections of sinks $s^{\ast}_k$ and $t^{\ast}_k$, collections of quasi-invariant sets $S_k$ 
and cycles $C_k$. The cycle $C_k$  is the cycle $C$ (including all their sub-cycles) containing sink $s^{\ast}_k$ that had been most recently created 
before the optimal W-graph $g^{\ast}_k$ was built (and hence the vertex $s^{\ast}_k$ lost its {} sink status). 
If the sink $s^{\ast}_k$ has not  been a part of any  cycle created before $g^{\ast}_k$ is built, then we set $C_k=\{s^{\ast}_k\}$.
The collection of cycles $C_k$ is a subset of the set of Freidlin's cycles \cite{freidlin-cycles,freidlin-physicad,f-w}.
Further, we, abusing notations, will denote the set of vertices in the cycle $C$ also by $C$.

\subsection{A pseudocode}
\label{sec:pse}
The  algorithm is summarized in the pseudocode below. 

 \vspace{3mm}
 \noindent
{\bf Algorithm 1.} \emph{A single-sweep algorithm constructing the hierarchy of the optimal W-graphs.}

\noindent
 {\bf Input}:~{For each vertex $i\in S$, list the set of outgoing arcs with  their weights $U_{ij}$ and pre-factors $a_{ij}$};\\
{\bf Output}:~{The collections of numbers $A_k$, $\Delta_k$, the collections of sets $S_k$ and $C_k$, the disappearing sinks $s^{\ast}_k$,
and the absorbing sinks $t^{\ast}_k$, the collection of exit arcs $(p^{\ast}_k\rightarrow q^{\ast}_k)$, and the collection of the
optimal W-graphs $g^{\ast}_k$ for $n-1\le k\le 1$};\\
~~$//$~~{\tt  Initialization}\\
{\bf 1.} Denote the set of outgoing arcs from vertex $i$ by $B_i$. 
Sort the arcs in each set  $B_i$ according to their weights $U_{ij}$ in the increasing order;\\
{\bf 2.} For each vertex $i\in S$, denote the outgoing arc of minimal weight by \verb|min_arc|$(i)$. Remove it from the set $B_i$.
Create the set of minimum outgoing arcs $M=\bigcup_{i\in S}$\verb|min_arc|$(i)$. Set $C(i) = \{i\}$; $G^{\ast}=\emptyset$;\\
{\bf 3.} Set $k = n-1$;\\
~~$//$~~{\tt The main cycle}\\
{\bf while} {\{$|M| > 1$\}}{ \\
  \indent {\bf 4.} Remove the minimum weight arc $(x\rightarrow y)$ from the set $M$ {and add it to the set $G^{\ast}$};\\
\indent {\bf if} { \{$x$ and $y$ belong to different connected components of $g^{\ast}_{k+1}$\}}  \\
{ 
 \indent\hspace{5mm}  {\bf 5.} Set $(p^{\ast}_k\rightarrow q^{\ast}_k) = (x\rightarrow y)$. 
 The sinks $s_{k}^{\ast}$ and $t_{k}^{\ast}$ are the sinks of the connected components of $g^{\ast}_{k+1}$ 
 containing the vertices $x$ and $y$ respectively. The set $S_k$ is the set of vertices in the connected component containing $s^{\ast}_{k}$. 
 Define $C_k =C(s^{\ast}_k)$;\\
  \indent\hspace{5mm} {\bf 6.} Set $\Delta_k = U_{xy}$ and $A_k = a_{xy}$;\\
  \indent\hspace{5mm} {\bf 7.} Find the connected component of the optimal W-graph $g^{\ast}_k$ 
  that results from merging of two connected components of $g^{\ast}_{k+1}$
  containing vertices $x$ and $y$ as follows. Starting from the sink of the connected component of 
  $g_{k+1}^{\ast}$ containing the vertex $y$, trace incoming arcs  
  from the set $G^{\ast}$  backwards until all vertices of this connected component are reached and every vertex except for the sink has a single outgoing arc;\\
  \indent\hspace{5mm} {\bf 8.} Set $k=k-1$;\\
   }
\indent  {\bf else} \\
 \indent\hspace{5mm}  { ~~$//$~~{\tt a cycle is created};\\
  \indent\hspace{5mm} {\bf 9.} Detect the cycle $C$;\\
  \indent\hspace{5mm} {\bf 10.} For each vertex $i\in C${,} except for $i=x${,} update the weights  and the pre-factors of the outgoing arcs  in the set $B_i$
   according to Eq. \eqref{update_rule}; \\
  \indent\hspace{5mm}  {\bf 11.} Merge sets $B_i$ for all $i\in C${, i.e., define} $B: = \bigcup_{i\in C}B_i$. Set $C' = \bigcup_{i\in C} C(i)$; \\
  \indent\hspace{5mm} {\bf while} { \{the vertices $r$ and $t$, where $(r\rightarrow t):= \arg\min_{(p\rightarrow q)\in B} U_{pg}$, belong to $C'$\}} \\  
     \indent\hspace{10mm} {\bf 12.} Discard the arc $(r\rightarrow t)$ from $B$;\\
 \indent\hspace{5mm}  {\bf end while}\\
  \indent\hspace{5mm}  {\bf 13.} Set $B_i=B$ and \verb|min_arc|$(i) = \arg\min_{(p\rightarrow q)\in B} U_{pg}$ for  all $i\in C'$. Set $C(i) = C'$ for all $i\in C'$;    \\
  \indent\hspace{5mm} {\bf 14.} Delete the minimum outgoing arc from $B$ and add it to $M$;\\
}
   \indent  {\bf end if - else}\\
    {\bf end while}

Algorithm 1 was motivated by $(i)$ Kruskal's algorithm \cite{kruskal,amo} for finding a minimum spanning tree in an undirected weighted graph,
and $(ii)$ Chu - Liu/Edmonds' algorithm \cite{chu-liu,edmonds} designed 
for finding
a spanning arborescence of a minimum weight with the root at the prescribed vertex 
(i.e., the directed analog of the minimum spanning tree).

\subsection{An illustrative example}
\begin{figure}[htbp]
\begin{center}
\includegraphics[width=\textwidth]{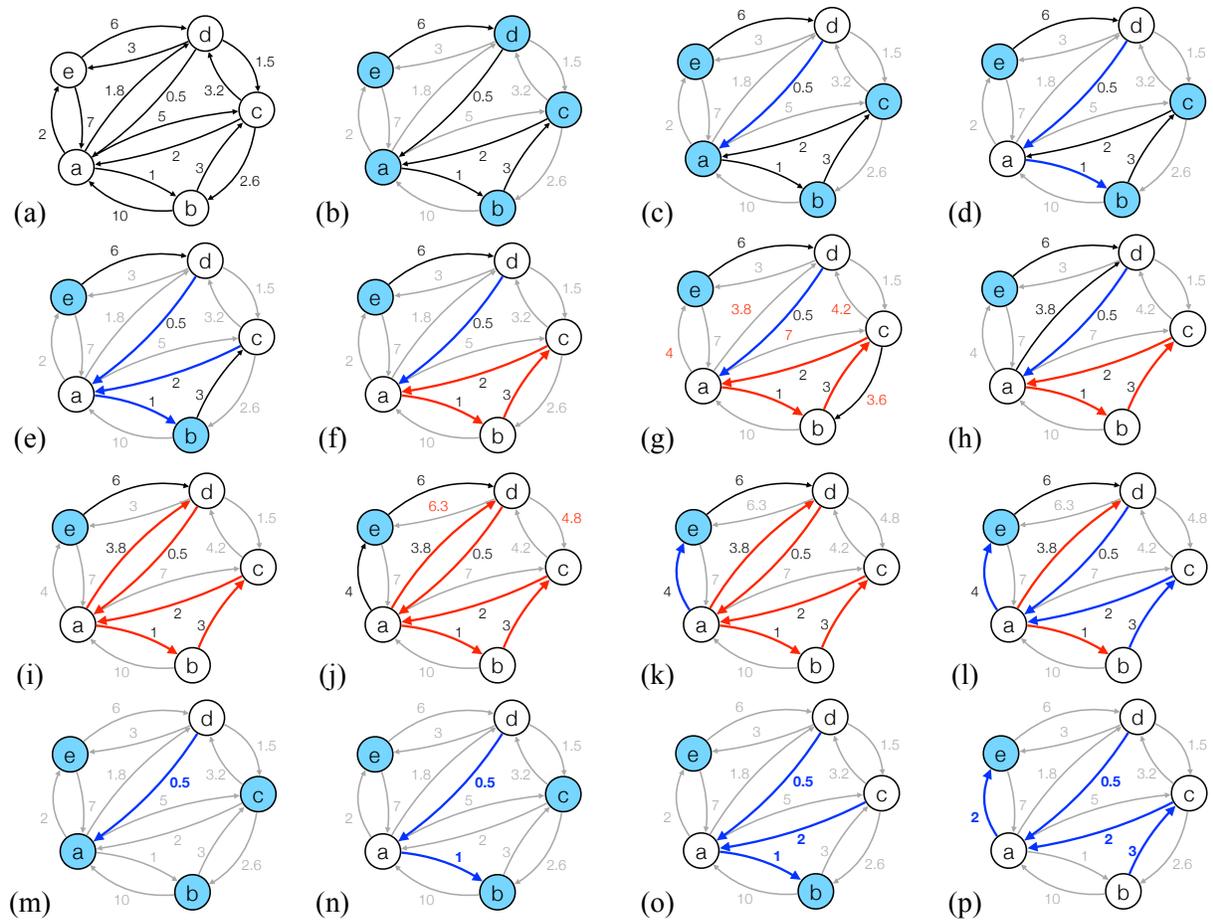}
\caption{An example demonstrating how Algorithm 1 works. 
}
\label{fig:OWGex}
\end{center}
\end{figure}
An example  in Fig. \ref{fig:OWGex} demonstrates how Algorithm 1 works.
The given continuous-time Markov chain is shown in Fig. \ref{fig:OWGex}(a). For reader's convenience, we have labeled statements
in Algorithm 1 with bold arabic numbers. We will refer to them as we work out the example.
{\bf 1:}  {First, we sort} the sets of outgoing arcs for each vertex { and} get:
\begin{align*}
B_a &= \{(a\rightarrow b; 1),~(a\rightarrow d; 1.8),~(a\rightarrow e;2),~(a\rightarrow c;5)\};\\
B_b &= \{(b\rightarrow c; 3),~(b\rightarrow a; 10)\};\\
B_c &= \{(c\rightarrow a; 2),~(c\rightarrow b; 2.6),~(c\rightarrow d;3.2)\};\\
B_d &= \{(d\rightarrow a; 0.5),~(d\rightarrow c; 1.5),~(d\rightarrow e;3)\};\\
B_e &= \{(e\rightarrow d; 6),~(e\rightarrow a; 7)\}.
\end{align*}
{\bf 2:} Next, we remove the minimal weight arc from each set $B_i$, $i\in\{a,b,c,d,e\}$, form the set $M$
out of them and sort the arcs in $M$:
\begin{align*}
B_a &= \{(a\rightarrow d; 1.8),~(a\rightarrow e;2),~(a\rightarrow c;5)\};\\
B_b &= \{(b\rightarrow a; 10)\};\\
B_c &= \{(c\rightarrow b; 2.6),~(c\rightarrow d;3.2)\};\\
B_d &= \{(d\rightarrow c; 1.5),~(d\rightarrow e;3)\};\\
B_e &= \{(e\rightarrow a; 7)\};\\
M & =\{(d\rightarrow a; 0.5),~(a\rightarrow b; 1),~(c\rightarrow a; 2),~(b\rightarrow c; 3),~(e\rightarrow d; 6)\};\\
C(a)& = \{a\},~ C(b) = \{b\} ,~ C(c) = \{c\},~ C(d) = \{d\},~ C(e) = \{e\}; 
\end{align*}
The arcs constituting the set $M$ are the thick black ones in Fig. \ref{fig:OWGex}(b).
{\bf 3:} Set $k = 4$. 

Now the main cycle starts. 
{\bf 4:} The minimal weight arc $(d\rightarrow a; 0.5)$ is removed from $M$ and added to  $G^{\ast}$.
Since $d$ and $a$ belong to different connected components of $g^{\ast}_5$ 
(containing all 5 sinks and no arcs), we execute statements {\bf 5 - \bf 8}
of {\bf if}. 
{\bf 5:}
$$
(p^{\ast}_4,q^{\ast}_4) = (d\rightarrow a);
\quad s^{\ast}_4 = d;\quad t^{\ast}_4 = a;\quad S_4 =C_4= \{d\}.
$$
{\bf 6:}
$$
\Delta_{4} = 0.5;\quad A_4 = a_{da}.
$$
{\bf 7:} $g^{\ast}_{4}$ contains the single arc $(d\rightarrow a)$ (shown in blue in Fig. \ref{fig:OWGex}(c)).
{\bf 8:} Set $k = 3$ and return to the beginning of the main {\bf while}-cycle.

{\bf 4:} The next minimal weight arc removed from $M$ and added to $G^{\ast}$ is $(a\rightarrow b; 1)$. 
Its endpoints $a$ and $b$ belong to different connected components of $g^{\ast}_4$. Hence we 
execute the statements of {\bf if}. 
 {\bf 5:}
 $$
(p^{\ast}_3,q^{\ast}_3) = (a\rightarrow b);\quad
\Delta_{3} = 1;\quad t^{\ast}_3 = b,\quad S_3 = \{a,d\};\quad C_3=\{a\}.
$$
{\bf 6:}
$$
A_3 = a_{ab};\quad s^{\ast}_3 = a.
$$
{\bf 7:} $g^{\ast}_3$ has two arcs
$(d\rightarrow a; 0.5)$ and $(a\rightarrow b; 1)$ shown in blue in Fig. \ref{fig:OWGex}(d). 
{\bf 8:} Set $k = 2$ and return to the beginning of the main {\bf while}-cycle.

{\bf 4:} The next minimal weight arc removed from $M$  and added to $G^{\ast}$ is $(c\rightarrow a; 2)$. 
Its endpoints belong to different connected components of $g^{\ast}_3$,
hence we execute the statements  of {\bf if}.
{\bf 5 :}
$$
(p^{\ast}_2,q^{\ast}_2) = (c\rightarrow a);\quad
 s^{\ast}_2 = c;\quad t^{\ast}_2 = b;\quad S_2=C_2 = \{c\}.
$$
{\bf 6:}
$$
\Delta_{2} = 2;\quad A_2 = a_{ca}.
$$
{\bf 7:} $g^{\ast}_2$ has three arcs: $(d\rightarrow a; 0.5)$, $(a\rightarrow b; 1)$, and  $(c\rightarrow a; 2)$ shown in blue in Fig. \ref{fig:OWGex}(e). 
{\bf 8:} Set $k = 1$ and return to the beginning of the main {\bf while}-cycle.

{\bf 4:} The next minimal weight arc removed from $M$ and added to $G^{\ast}$ is $(b\rightarrow c; 3)$. 
Its endpoints belong to the same connected component of $g^{\ast}_2$.
Hence we execute the statements of {\bf else}.
{\bf 9:} We detect the cycle
$$
C = \{(b\rightarrow c),~(c\rightarrow a),~(a\rightarrow b)\}
$$ 
shown in Fig. \ref{fig:OWGex}(f) in red.
{\bf 10:} Then we increase the arc weights in the sets $B_a$ and $B_c$ using the update rule \eqref{update_rule}  
by $3-1 = 2$ and $3-2 = 1$ respectively:
\begin{align*}
B_a &= \{(a\rightarrow d; 3.8),~(a\rightarrow e;4),~(a\rightarrow c;7)\};\\
B_c &= \{(c\rightarrow b; 3.6),~(c\rightarrow d;4.2)\}.
\end{align*}
The updated arc weights are shown in Fig. \ref{fig:OWGex}(g) in red.
The pre-factors of arcs in $B_a$ and $B_c$ are multiplied by $a_{bc}/a_{ab}$ and $a_{bc}/a_{ca}$ respectively.
{\bf 11:} Merge the sets $B_a$, $B_b$ and $B_c$ and sort the arcs in the resulting set:
$$
B_{abc}=\{(c\rightarrow b; 3.6),~(a\rightarrow d; 3.8),~(a\rightarrow e;4),~(c\rightarrow d;4.2),~(a\rightarrow c;7),~(b\rightarrow a; 10)\}.
$$
Define the cycle $C' = C(a)\cup C(b)\cup C(c) = \{a,b,c\}$.
{\bf 12:}  {Discard} the minimal weight arc $(c\rightarrow b; 3.6)$ from $B_{abc}$  because its endpoints belong to the same cycle $C'$. 
{\bf 13:} {Set} $B_a = B_b = B_c = B_{abc}$ and $C(a)=C(b)=C(c)=C'=\{a,b,c\}$.
{\bf 14:} The next minimal weight arc in $B_{abc}$ is $(a\rightarrow d; 3.8)$. {Remove} it from $B_{abc}$ and add to $M$ (Fig. \ref{fig:OWGex}(h)). 
At this point, $M$ is:
$$
M=\{(a\rightarrow d; 3.8),~(e\rightarrow d; 6)\}.
$$
Now, we return to the beginning of the main {\bf while}-cycle. 

{\bf 4:} {Remove} the minimal weight arc $(d\rightarrow a; 3.8)$ from $M$ and add it to $G^{\ast}$. 
Its endpoints $d$ and $a$ belong to the same connected component of $g^{\ast}_2$. Hence we execute the statements of {\bf else}.
We detect the cycle (Fig. \ref{fig:OWGex}(i))
$$
C= \{(a\rightarrow d),~(d\rightarrow a)\}.
$$ 
{\bf 10.} According to the update rule \eqref{update_rule},  the arc weights in the set $B_d$ are increased
by $3.8-0.5 = 3.3$:
$$
B_d = \{(d\rightarrow c; 4.8),~(d\rightarrow e;6.3)\},
$$
and their pre-factors are multiplied by the factor $a_{ad}/a_{da}$.
The updated weights are shown in Fig. \ref{fig:OWGex}(j) in red.
 {\bf 11:} Merge the sets $B_d$ and $B_{abc}$ and sort the arcs in the resulting set:
$$
B_{abcd} = \{(a\rightarrow e;4),~(c\rightarrow d;4.2),~(d\rightarrow c; 4.8),~(d\rightarrow e;6.3),~(a\rightarrow c;7),~(b\rightarrow a; 10)\}.
$$
Define the cycle $C' = C(a)\cup C(d) = \{a,b,c\}\cup \{d\} = \{a,b,c,d\}$.  
The endpoint $e$ of the minimal weight arc in $B_{abcd}$ 
$ (a\rightarrow e;4)$ does not belong to $C'$, hence we skip {\bf 12}. 
 {\bf 13:} {Set} $B_i = B_{abcd}$, \verb|min_arc|$(i) = (a\rightarrow e)$, and $C(i) = \{a,b,c,d\}$ for $i=a,b,c,d$.
 {\bf 14:} The minimal weight arc $(a\rightarrow e;4)$ is removed from $B_{abcd}$ and added to $M$.
Now, we return to the beginning of the main {\bf while}-cycle. 

{ {\bf 4:} }We remove the minimal weight arc $(a\rightarrow e;4)$ from $M$  and add it to $G^{\ast}$ (Fig. \ref{fig:OWGex}(k)).
Its endpoints $a$ and $e$ belong to different connected components of $g^{\ast}_2$, hence we execute the statements of {\bf if}.
{\bf 5:} 
$$
(p^{\ast}_1,q^{\ast}_1) = (a\rightarrow e);
\quad s^{\ast}_1 = b;\quad t^{\ast}_1 = e,\quad S_1=C_1 = \{a,b,c,d\}.
$$
{\bf 6:}
$$
\Delta_{1} = 4;\quad A_1 = \frac{a_{ae}a_{bc}}{a_{ab}}.
$$
{\bf 7:} The optimal W-graph $g^{\ast}_1$ is found by tracing the arcs  
from the set $G^{\ast}$ backwards starting from the sink $e = t^{\ast}_1$ so that each encountered vertex has
only one outgoing arc (Fig. \ref{fig:OWGex}(l)). The arcs in $G^{\ast}$ are  blue and red.  The arcs in $g^{\ast}_1$ are blue.
{\bf 8:} Set $k = 0$. Now, $|M| =|\{(e\rightarrow d; 6)\}| = 1$. Therefore, the main {\bf while}-cycle is {terminated}.
The found optimal W-graphs $g^{\ast}_4 - g^{\ast}_1$ are shown in Figs. \ref{fig:OWGex}(m) - (p) respectively.  

From the output of Algorithm 1, we obtain the following asymptotic approximations for nonzero eigenvalues:
\begin{equation}
\lambda_4\approx a_{da}e^{-0.5/T},\quad 
\lambda_3\approx a_{ab}e^{-1/T},\quad 
\lambda_2\approx a_{ca}e^{-2/T},\quad 
\lambda_1\approx  \frac{a_{ae}a_{bc}}{a_{ab}}e^{-4/T}.
\end{equation}

\subsection{Implementation, cost,  and performance}
We have implemented Algorithm 1 both in Matlab and in C. 
Our C code  uses pointers and such data structures as binary trees and link lists.
All arcs are stored as structures {containing} three pointers to structures of the same type: parent, left child, and right child.
The arcs { in the sets $B_i$, $i\in S$, and $M$} are organized in binary trees automatically sorted using the heap sort \cite{amo}.
{As a result, removing or adding} an arc from/to any of these sets { is reduced} 
to reshuffling the corresponding pointers to parents and children.
{Link lists are used to keep track  of  connected components of optimal W-graphs and  Freidlin's cycles  
in a similar manner as  in the description of  Kruskal's algorithm in \cite{amo}. }

The computational cost of Algorithm 1 depends on the number vertices { and} the distribution{s} of arcs and their weights.
{Let  a network have $n$ vertices and the maximal out-degree be $d$}.
The cost of the initialization is due to building  $n$ binary trees $B_i$, 
the removal of their roots{,} and building   the main binary tree $M$
out of them. It adds up to
\begin{equation}
\label{cost1}
O(nd\log d ) +O( n\log d) +O( n\log n).
\end{equation}
The computational cost of the main while-cycle depends on the complexity on the distribution of arcs and their  weights.
In the best case scenario, i.e., when no cycle is created and hence the else-statements are never executed, 
the cost is merely due to arc removals from the main tree $M$. It is 
\begin{equation}
\label{cost2}
O( n\log n).
\end{equation}
In the worst case scenario, $n-1$ cycles can be created. Then the computational cost will be dominated by 
the cost of merging  the trees $B_i$. It will be at most
\begin{equation}
\label{cost3}
\sum_{l=d}^{nd}O(l\log l) =O((nd)^2\log (nd)).
\end{equation}

The performance of Algorithm 1 has been tested on the networks LJ$_{38}$ and LJ$_{75}$ created by Wales' group \cite{web}.
The LJ$_{38}$ network \cite{pathsample} contains 71,887 vertices and 239,706 arcs. 
The CPU time of Algorithm 1 applied to LJ$_{38}$ on a single core of an iMac desktop is 30 seconds. 
The number of created cycles  is 50,226.
The LJ$_{75}$ network contains 169,523 vertices and 441,016 arcs. 
The CPU time  of Algorithm 1 applied to LJ$_{75}$ is 632 seconds. The number of  created cycles  is 153,164.


\section{An application to LJ$_{75}$}
\label{sec:LJ75}
In this Section, we will present 
our analysis of the LJ$_{75}$ network. To make sure, it is time-reversible.
For the purposes of the spectral analysis, we will restrict our attention to the
largest connected component of   LJ$_{75}$ 
containing 169,523 vertices.
We will refer to the { vertices by the indices of the corresponding local minima }
in the full dataset of 593,320 minima. 
A detailed description of Lennard-Jones clusters and, in particular,  of LJ$_{75}$, can be found in \cite{wales_landscapes,wales_book}.
Here we just remind that the energy landscape of LJ$_{75}$ has a double-funnel structure. 
Minimum  1 (vertex 1) in the LJ$_{75}$ network has the lowest energy and lies at the bottom of the deep and narrow funnel. 
It corresponds to the configuration shown in Fig. \ref{fig1}(a), the Marks decahedron with the point group $D_{5h}$.
There are several  local minima at the bottom of the wide and shallower icosahedral funnel. The one with the second lowest energy
is minimum  92 (vertex 92) shown in Fig. \ref{fig1}(b). 


The pairwise transition rates  in the networks representing energy landscapes of Lennard-Jones clusters 
can be modeled using a harmonic approximation as it is done in \cite{wales0}.
In this case, the off-diagonal entries of the generator matrix are
\begin{equation}
\label{LJrates}
L_{ij} = \frac{O_i\left(\Pi_{+}\nu_i\right)^{1/2}}{O_{ij}\left(\Pi_+\nu_{ij}\right)^{1/2}}e^{-(V_{ij}-V_i)/T},
\end{equation}
where { the subscripts $i$  and $ij$ refer to  the minimum $i$
and the transition state $(ij)$ separating the minima $i$ and $j$ respectively. 
$O_i$ and $O_{ij}$ are the point group orders  of $i$ and $(ij)$, 
$\Pi_{+}\nu_i$ and $\Pi_{+}\nu_{ij}$ are the products of the positive eigenvalues of the Hessian matrices at $i$ and $(ij)$,
 and $V_{i}$ and $V_{ij}$
are the values of the potential at $i$ and $(ij)$ respectively.}
The invariant probability distribution is given by
\begin{equation}
\label{LJinv}
\pi_i= \frac{1}{Z}\frac{e^{-V_i/T}}{O_i\left(\Pi_{+}\nu_i\right)^{1/2}},\quad \text{where}\quad
Z=\sum_{i}\frac{e^{-V_i/T}}{O_i\left(\Pi_{+}\nu_i\right)^{1/2}}.
\end{equation}

\subsection{Zero-temperature asymptotic analysis}
\label{sec:LJ75zero}
\subsubsection{Asymptotic estimates for eigenvalues of LJ$_{75}$}
Eq. \eqref{eq8} implies that the asymptotic estimates for the eigenvalues of generator matrices with 
off-diagonal entries of the form of Eq. \eqref{LJrates} are given by
\begin{equation}
\label{LJevals}
\lambda_k = \frac{O_{s^{\ast}_k}\left(\Pi_{+}\nu_{s^{\ast}_k}\right)^{1/2}}
{O_{p^{\ast}_kq^{\ast}_k}\left(\Pi_+\nu_{p^{\ast}_kq^{\ast}_k}\right)^{1/2}}e^{-(V_{p^{\ast}_kq^{\ast}_k}-V_{s^{\ast}_k})/T}.
\end{equation}
The collection of numbers $\Delta_k = V_{p^{\ast}_kq^{\ast}_k}-V_{s^{\ast}_k}$ for the LJ$_{75}$ network is shown in 
Fig. \ref{fig:LJ75deltas}.
\begin{figure}[htbp]
\begin{center}
\includegraphics[width = 0.6\textwidth]{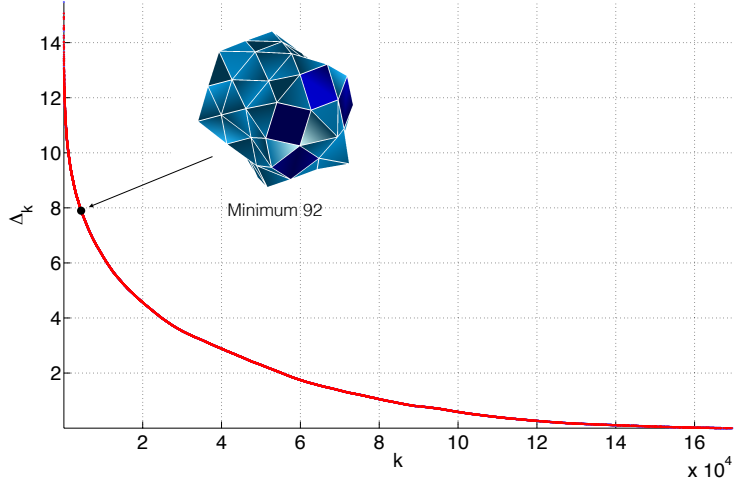}
\caption{The exponential factors $\Delta_k$, $1\le k\le 169522$, determining the asymptotic
estimates for the eigenvalues of the generator matrix of LJ$_{75}$ according to Eq. \eqref{LJevals}. }
\label{fig:LJ75deltas}
\end{center}
\end{figure}
As one can see, there are no  notable gaps separating the exponential factors $\Delta_k$'s. Hence, there is no 
appreciable scale separation for the transition processes going on in LJ$_{75}$.
As a result, one cannot find any collection of metastable points (a subset of states),
with respect to which this Markov chain would be metastable according to the definition in \cite{bovier2002,bovier1}.
A similar phenomenon was observed in the LJ$_{38}$ network \cite{cspec1}.

The eigenvalue corresponding to the escape process from the icosahedral funnel 
is the one computed at that step of Algorithm 1 { at which} the vertex 92 
ceases to be a sink. This happens at $k=4395$. 
Therefore, minimum 92 is the sink $s^{\ast}_{4395}$.
The corresponding sink $t^{\ast}_{4395}$ is minimum 1. Hence, the escape from the icosahedral
funnel brings the cluster into the Marks decahedron funnel with probability close to one at small enough temperatures.
The pre-factor and exponent  for $k=4395$ found by Algorithm 1 are 
$A_{4395} = 147.2$ and $\Delta_{4395} = 7.897$. 
Hence, {at low temperatures, the theoretical escape rate from the icosahedral funnel
is given by }
\begin{equation}
\label{rate_theory}
\lambda_{4392}(T) \approx 147.2e^{-7.897/T}.
\end{equation}
We will verify this prediction in Section \ref{sec:LJ75finite}.
The arc $p^{\ast}_{4395}\rightarrow q^{\ast}_{4395}$ is found to be $(25811\rightarrow 73992)$.
The quasi-invariant set $S_{4395}$ contains 92883 states.

\subsubsection{The asymptotic zero-temperature path}
The asymptotic zero-temperature path (the MinMax path) \cite{cam1} connecting minima 92 and 1 can be extracted as the unique path 
from vertex $s^{\ast}_{4395} = 92$ to  vertex $t^{\ast}_{4395} = 1$ in the optimal W-graph $g^{\ast}_{4395}$.
The potential energy along this path relative to the energy at minimum 1 is plotted in Fig. \ref{fig:path0}{.}
\begin{figure}[htbp]
\begin{center}
\includegraphics[width = 0.6\textwidth]{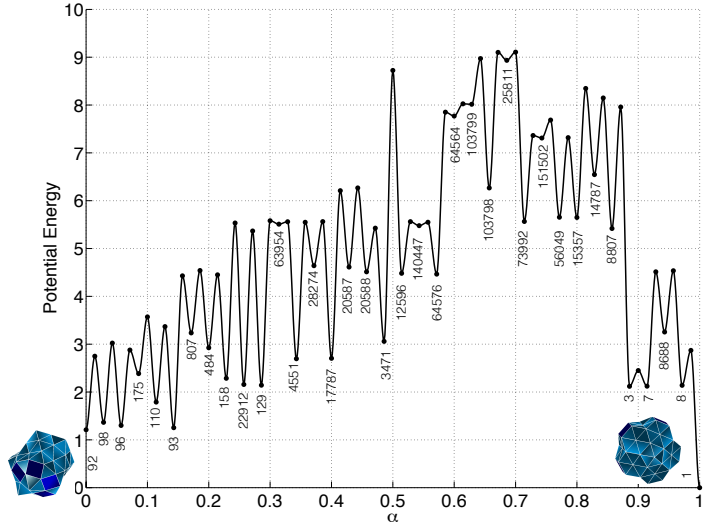}
\caption{The energy along the asymptotic zero-temperature path (the MinMax path) connecting the two lowest minima 1 and 92 of  LJ$_{75}$.
The energy at the global minimum (minimum 1) is assumed to be zero. 
The numbers next to each minimum indicate the indices of the corresponding local minima in 
D. Wales's { full} dataset.}
\label{fig:path0}
\end{center}
\end{figure}

\subsubsection{The decomposition in the maximal quasi-invariant sets}
A quasi-invariant set $S_k$ is called maximal if it is not a subset of any other quasi-invariant set $S_l$ for $0<l<k$.
In other words, if  $t^{\ast}_k = s^{\ast}_0$, i.e., the connected component with the sink $s^{\ast}_k$
 is absorbed by the connected component with sink $s^{\ast}_0$ (the only sink of $g^{\ast}_1$).
The set of states of the LJ$_{75}$ network can be decomposed into the union of  sink $s^{\ast}_0$ {(minimum 1)}
and 15692 maximal quasi-invariant sets. 
The largest maximal quasi-invariant set is  $S_{4395}$ with sink $s^{\ast}_{4395}= 92$. 
All maximal quasi-invariant sets with more than 1000 local minima are listed in Table 1. The numbers $|C_k|$ are the numbers of
local minima (vertices) in the sets  $C_k$ constructed by Algorithm 1. 
The sets $C_k$ consist of all vertices separated from the sink $s^{\ast}_k$ by energy
barriers less than $\Delta_k$. They are the maximal Freidlin's cycles containing minimum $s^{\ast}_k$
and not containing any  local minimum with energy 
less than $V_{s^{\ast}_k}$ \cite{freidlin-cycles, freidlin-physicad,f-w,cam1,cspec1}.
\begin{table}[htdp]
\caption{The maximal quasi-invariant sets in LJ$_{75}$ containing over 1000 local minima.}
\begin{center}
\begin{tabular}{|c|c|c|c|c|c|}
\hline
$k$&$s^{\ast}_k$&$|S_k|$&$|C_k|$&$\Delta_k$\\
\hline
4395 & 92  & 92883 & 28032 & 7.897\\
47688 & 2  & 7141 & 27 & 2.418\\
71664 & 24  & 5026 & 52 & 1.308\\
30880 & 18438  & 2609 & 177 & 3.465 \\
41622 & 135496  & 2590 & 38 & 2.785\\
\hline
\end{tabular}
\end{center}
\label{table1}
\end{table}%

A diagram showing the mutual arrangement of the maximal quasi-invariant sets containing more that 100 local minima is shown in Fig. \ref{fig:s}.
Each shown maximal quasi-invariant set $S_k$
is represented as a circle whose area is proportional to the number of minima in {$S_k$}.
The black number next to each circle is the index of the local minimum  corresponding { to the} sink $s^{\ast}_k$.
The { corresponding arcs $(p^{\ast}_k\rightarrow q^{\ast}_k)$ are shown with the black arrows. The} dark red numbers next to 
them are the  exponential factors $\Delta_k$.  
\begin{figure}[htbp]
\begin{center}
\includegraphics[width = 0.6\textwidth]{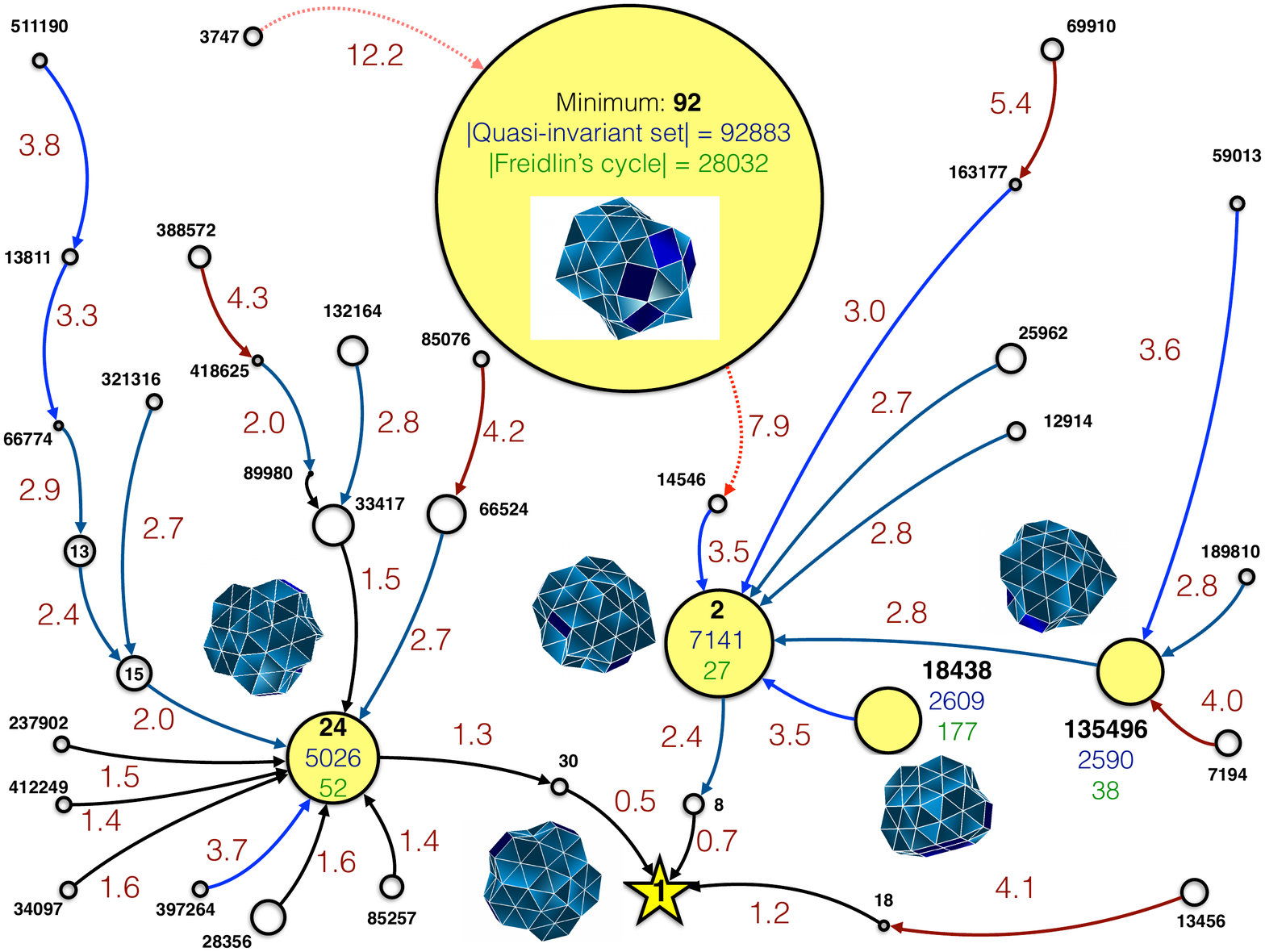}
\caption{The maximal quasi-invariant sets in LJ$_{75}$ containing over 100 local minima.
}
\label{fig:s}
\end{center}
\end{figure}

\subsubsection{Using bond-order parameters $Q_4$ and $Q_6$}
The set of local potential minima of LJ$_{75}$ can be mapped into a low-dimensional space of bond-order parameters 
introduced in \cite{cop1,cop2} in order to distinguish different types of atomic packings.
The bond-order parameters are defined by
$$
Q_l = \left[\frac{4\pi}{2l+1}\sum_{m=-l}^l|\langle{Y}_{lm}(\theta(\mathbf{r}_{ij}),\phi(\mathbf{r}_{ij})\rangle|^2\right]^{1/2},
$$ 
where ${Y}_{lm}(\theta(\mathbf{r}_{ij}),\phi(\mathbf{r}_{ij})$ are the spherical harmonics, 
$\theta$ and $\phi$ are the polar angle{s} of the radius-vector of the bond $\mathbf{r}_{ij}$, 
and the average 
 $\langle{Y}_{lm}(\theta(\mathbf{r}_{ij}),\phi(\mathbf{r}_{ij})\rangle$ is taken over all bonds in the cluster.
The bond-order parameters $(Q_4,Q_6)$ were used in e.g. \cite{wales38,neirotti,picciani,mfc75} 
in order to monitor  structural transitions  in {Lennard-Jones clusters}. 
Following \cite{wales38}, we set that  atoms $i$ and $j$ in the cluster form a bond  $\mathbf{r}_{ij}: = \mathbf{r}_i-\mathbf{r}_j$ if
$|\mathbf{r}_i-\mathbf{r}_j| < 1.391r^{\ast}$, where  $\mathbf{r}_i$ and $\mathbf{r}_j$ are the positions of atoms $i$ and $j$ 
respectively, and $r^{\ast} =2^{1/6}$ is the { minimizer} of the pair potential $V(r) = 4(r^{-12} -r^{-6})$.

In this Section, our goal is to investigate whether and how the bond-order parameters $(Q_4,Q_6)$
can be used for the study of the transition process between icosahedral and Marks decahedral configurations  in LJ$_{75}$.
The set of states belonging to the quasi-invariant set $S_{4395}$ with the sink $s^{\ast}_{4395}=92$
is shown with  the light { and dark} blue dots in Fig. \ref{fig:q}, while the rest of the states is shown with the pink {and} red dots. 
These sets of dots are all intermixed, hence it is  impossible to separate them in the $(Q_4,Q_6)$ {plane}.
However, let us turn our attention to Freidlin's cycles. 
{Freidlin's cycle $C_{4395}$, i.e., the set of local minima separated from minimum 92 by lower potential barriers than $\Delta_{4395}=7.897$,
is shown with dark blue dots in Fig. \ref{fig:q}.}
It consists of 28032
local minima. The larges{t} Freidlin's cycle containing minimum 1 and not containing minimum 92 consists of 15838 local minima.
It is shown with red dots. These sets of dots are intermixed only in a small region around $Q_4=0.015$ and $Q_6 = 0.17$. We will refer to {it as the}
transition region.
The  MinMax path is shown with the yellow piecewise linear curve. 
Comparing Figs. \ref{fig:path0} and \ref{fig:q}
we see that the MinMax path first wanders in the icosahedral corner ($Q_4 < 0.01$, $Q_6 < 0.03$) 
while passing through states 92 -- 129. Next, it switches to the transition region passing though states 63954, 4551 and 2874.
Then, it wanders in the transition region where it overcomes the highest potential barriers and passes through states 17787 -- 8807.
Finally, it jumps to state 3 lying in the Marks decahedral region ($Q_4\approx 0.03$ and $Q_6\approx 0.3$) 
and remains there while passing through states 7, 8688, 8{,} and reaching state 1.
Hence, despite not perfect, the coordinates $(Q_4,Q_6)$ can be used as reaction coordinates for the 
transition process between the two lowest minima of LJ$_{75}$.
\begin{figure}[htbp]
\begin{center}
\includegraphics[width = 0.6\textwidth]{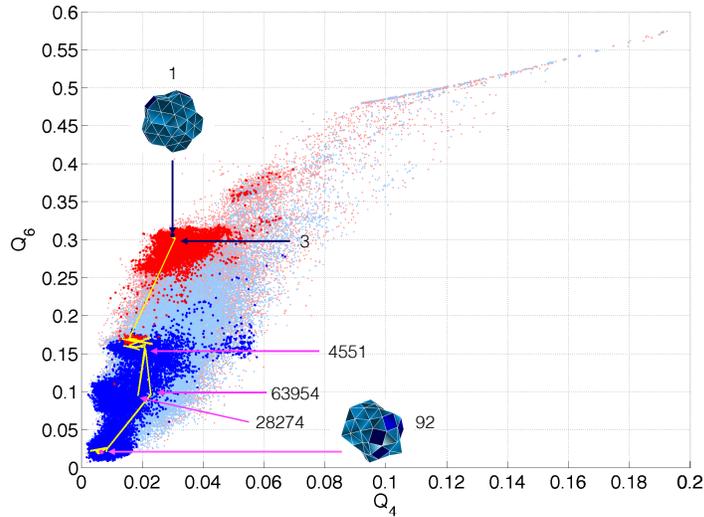}
\caption{Some results of the asymptotic spectral analysis of LJ$_{75}$ presented 
using bond-order parameters $(Q_4,Q_6)$.
Red dots: the largest Freidlin's cycle containing 
minimum 1 that does not contain minimum 92 (15838 local minima).
Dark blue dots: the largest Freidlin's cycle containing 
minimum 92 that does not contain minimum 1 (28032 local minima).
Light blue dots: the quasi-invariant set 
associated with minimum 92  (92883 local minima).
Pink dots:  the set of minima not contained 
in the quasi-invariant set 
associated with minimum 92  (76640 local minima).
Piecewise linear yellow curve: the asymptotic zero-temperature path (the MinMax path).
}
\label{fig:q}
\end{center}
\end{figure}

\subsection{Finite temperature analysis}
\label{sec:LJ75finite}
In this Section, we describe our finite temperature continuation technique
developed for computing the eigenvalue and the right eigenvector corresponding (at low temperatures)
to the transition process from the icosahedral
to the Marks decahedron funnel. 
Then we  present a quantitative description of this transition process using the corresponding eigencurrent
calculated from the right eigenvector.
We will refer to these eigenvalue, right eigenvector 
and eigencurrent as $\lambda(ICO - MARKS)$, $\phi(ICO - MARKS)$,  and $F(ICO - MARKS)$
respectively.

\subsubsection{Continuation strategy}
\label{sec:con}
We would like to obtain the eigenpair $(ICO - MARKS)$ at the temperature range $0< T \le T^{LJ75}_{sl} = 0.25$ 
(the melting point). 
A similar task was accomplished for LJ$_{38}$ in \cite{cspec2}. However, the continuation technique
used for LJ$_{38}$ turned out to be unsuccessful for LJ$_{75}$. The LJ$_{75}$ network is larger and more complex than LJ$_{38}$.
It contains wider range of potential barriers. The solid-solid transition temperature  $T_{ss}=0.08$ \cite{75solid-solid} where Marks decahedral states give place to
 icosahedral configurations is extremely low (see Section \ref{sec:intro}) in comparison with the 
 energetic barrier of 7.897 separating minimum 92 from minimum 1. For $T<0.17$, 
Rayleigh quotient iteration{s} (e.g. \cite{trefethen}) fail 
on LJ$_{75}$ due to issues associated with floating point arithmetic (produces NaN or fails to converge). 
For $T\ge 0.17$, the initial approximation for the right eigenvector 
(the indicator function of $S_{4395}$) is not close enough to the right eigenvector $\phi(ICO-MARKS)(T)$ at $T\ge 0.17$, 
so that Rayleigh quotient iterations converge to a wrong eigenpair. 
As a result, we have nothing to anchor on.

To tackle these difficulties, we deployed two remedies, lumping and truncation, {and used a verification method relying on the eigencurrent}.
The resulting continuation strategy  enabled us to obtain the eigenvalue $\lambda(ICO - MARKS)$
for the temperature range $0.05\le T\le 0.25$ (see Fig. \ref{fig:con}).

\subsubsection{Initial approximation and verification}
As we have mentioned in Section \ref{sec:sign}, if the continuous-time Markov chain is time-reversible, which is always the case
if it represents an energy landscape, its generator matrix $L$ is similar a symmetric matrix $L_{sym}=P^{1/2}LP^{-1/2}$ where 
$P=diag\{\pi_1,\ldots,\pi_n\}$. 
The matrix of eigenvectors of $L_{sym}$ is orthogonal and given by $P^{1/2}\Phi$, where $\Phi$ is the matrix of right eigenvectors of $L$ normalized so that 
$\Phi^TP\Phi = I$.
The Rayleigh quotient iteration is applied
to the matrix $L_{sym} = L_{sym}(T)$ ({note that  $L_{sym}$ depends on temperature}).
Suppose we want to compute the eigenvector  $v(T)$ of $L_{sym}(T)$  at temperatures $T_1<T_2<\ldots <T_m$
corresponding to the escape process
described by the $k$th eigenpair $(\lambda_k,\phi^k)$ of $L$ at temperatures close to zero.

The initial approximation for $v(T_l)$, $1\le l\le m$, can be taken to be $P^{1/2}(T_l)\phi_0^k $, where
$\phi_0^k$ is the indicator function of $S_k$ and $P^{1/2}(T_l)=diag\{\pi_1^{1/2}(T_l),\ldots,\pi_n^{1/2}(T_l)\}$.
Alternatively, suppose the Rayleigh quotient iteration has converged to the desired eigenpair some  $T_r$. 
Then initial approximations for $v(T_l)$, $1\le l\le m$, $l\neq r$, can be chosen to be $P^{1/2}(T_l)P^{-1/2}(T_{r})v(T_{r})$.

Suppose we have calculated an eigenpair $(\lambda, v)$ of $L_{sym}$ at temperature $T$.
Then $(\lambda,w=P^{-1/2}(T)v)$ is the eigenpair of $L(T)$.
Now we need to verify whether it corresponds to the 
$k$th eigenpair $(\lambda_k,\phi^k)$ of $L$ at temperatures close to zero.
We claim that  $(\lambda,w=P^{-1/2}(T)v)$ is the desired eigenpair if  significant {fractions} of the corresponding eigencurrent are
emitted at the state $s^{\ast}_k$ and absorbed at the state $t^{\ast}_k$. 
Recall that the total amount of eigencurrent $F$, corresponding to the eigenpair $(\lambda,w)$ of $L$, 
emitted or absorbed at state $i$, is given by
$$
\sum_{j\neq i} F_{ij} = e^{-\lambda t}\lambda \pi_iw_i.
$$
Therefore, if we sort the components for the vector $P(T)w$ in the decreasing order, 
the component $w_{s^{\ast}_k}$ must be the first or nearly the first,
while the component $w_{t^{\ast}_k}$ must be last  or among the last ones.

\subsubsection{Lumping and truncation}
\label{sec:lump}
As we have mentioned, for $T<0.17$, the Rayleigh quotient iteration fails to converge
when we try to calculate the $ICO - MARKS$ eigenpair, while for $T\ge 0.17$, 
the iterations converge to wrong eigenpairs. 
If we manage to obtain a better initial approximation to the desired eigenpair, we can hope to calculate it 
for $0.17 \le T\le 0.25$. At lower temperatures, only {approximate eigenvalues} can
be calculated using a modified generator matrix $\tilde{L}$ instead of   $L$. 

The Rayleigh quotient iteration fails to converge for $T<0.17$ due to large range of orders of magnitude of its entries.  
{The largest } entries of $L$ can be eliminated by lumping together states in
each connected component of the optimal W-graph $g^{\ast}_k$ where  $k$ is the smallest number such that 
{$\Delta_k\le\Delta_{\min}$, a user-supplied threshold.}
{The smallest} entries of $L$ can be removed by capping the allowed values of the potential at states and edges by some user-supplied 
$V_{\max}$. Lumping and truncation are illustrated in Figs. \ref{fig:lump} (a) and (b) respectively.
\begin{figure}[htbp]
\begin{center}
\centerline{
(a)\includegraphics[width = 0.45\textwidth]{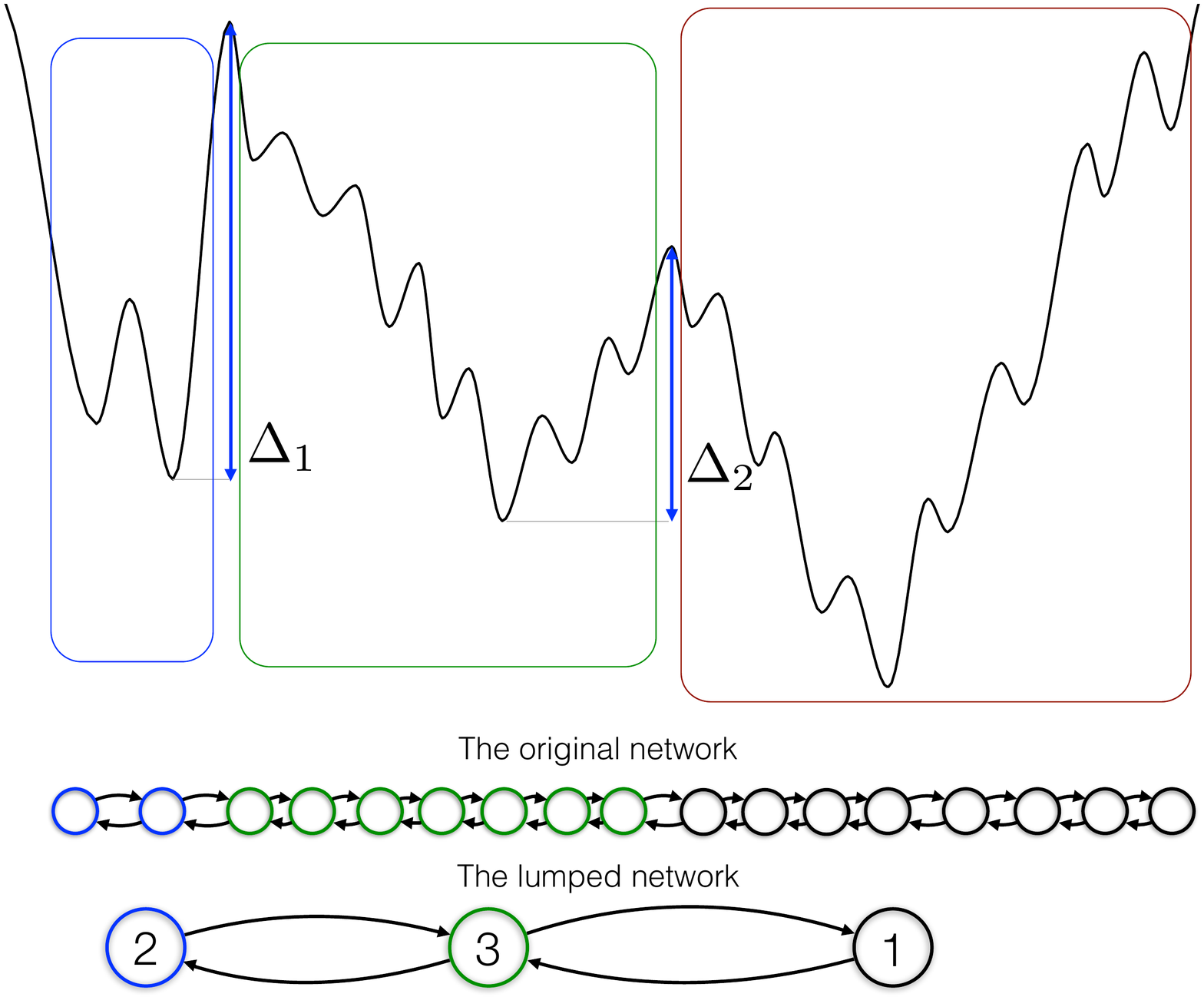}
(b)\includegraphics[width=0.45\textwidth]{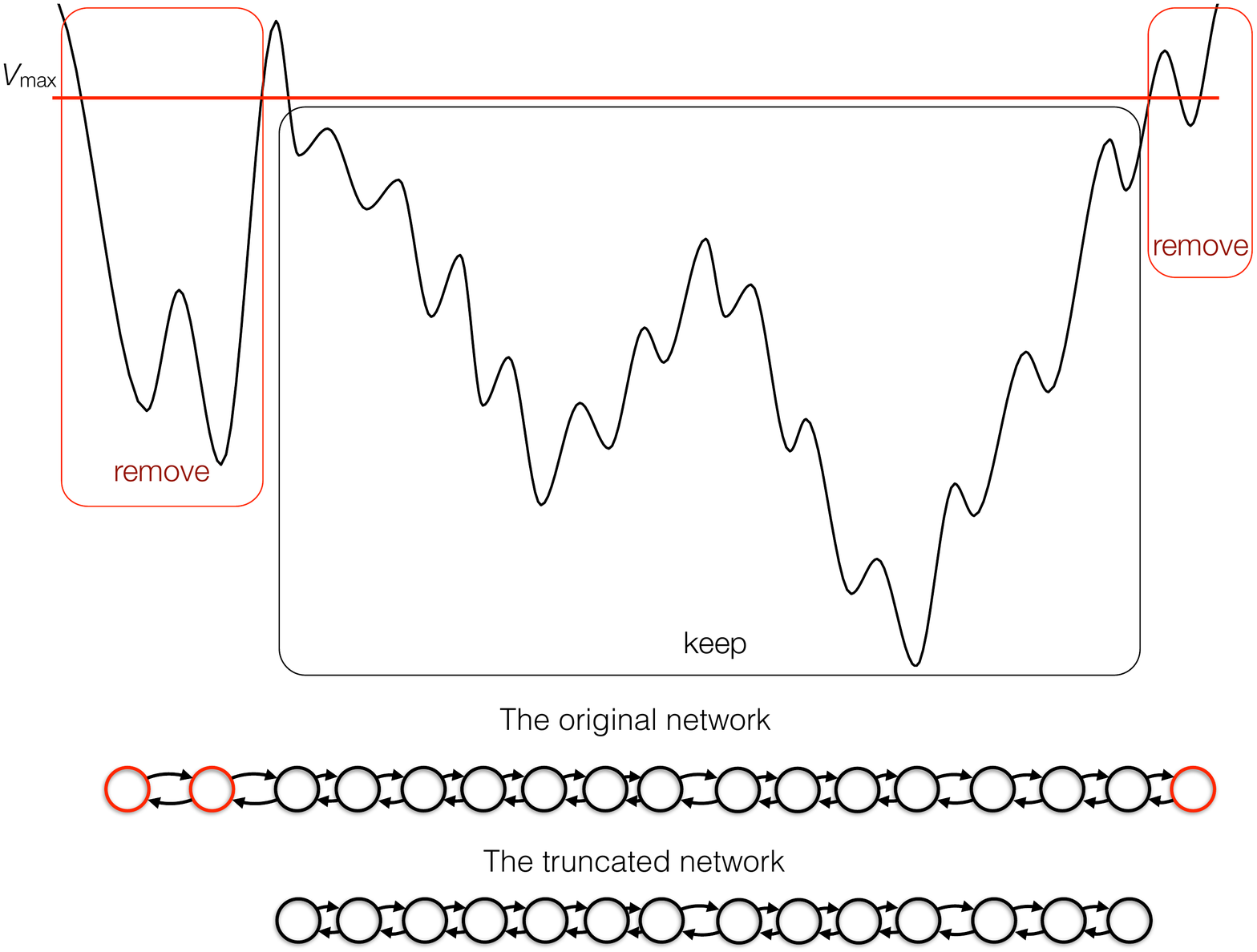}
}
\caption{(a) Lumping of the network using $\Delta_{\min} = \Delta_2$.
(b) Truncation of the network.
}
\label{fig:lump}
\end{center}
\end{figure}

The pairwise transition rates in the lumped network are calculated as follows. Let the optimal W-graph $g^{\ast}_k$ consist of
connected components $\Gamma_1$, ..., $\Gamma_k$.  Then the lumped network has $k$ states. 
The off-diagonal entries of its generator matrix $\tilde{L}$ are defined according to the formula
\begin{equation}
\label{lump_rate}
\tilde{L}_{lr} = \sum_{i\in \Gamma_l}\frac{\pi_i}{\sum_{i'\in\Gamma_l}\pi_{i'}}\sum_{j\in\Gamma_r}L_{ij},\quad l\neq r,
\quad \tilde{L}_{ll} = -\sum_{r\neq l}\tilde{L}_{lr},\quad 1\le j,r\le k.
\end{equation}
Eq. \eqref{lump_rate} says that the transition rate from
the connected component $\Gamma_l$ to the connected component $\Gamma_r$
is the sum of transition rates $L_{ij}$ along arcs $i\rightarrow j$ where $i\in \Gamma_l$ and $j\in\Gamma_r$ 
multiplied by the { conditional } probabilities
$\pi_i/\left(\sum_{i'\in\Gamma_l}\pi_{i'}\right)$
to find the system at state $i$  { provided that it is in }$\Gamma_l$.
The diagonal entries of $\tilde{L}$ are defined so that its row sums are zero.
 
One can check \cite{hs1,hs2} that the $k\times k$ generator matrix $\tilde{L}$ 
of the lumped network is obtained from the $n\times n$ generator matrix $L$ of the original network by multiplying $L$ by a $k\times n$ matrix $A$
from the left and by a $n\times k$ matrix $B$ from the right
\begin{equation}
\label{lump_mat}
\begin{array}{ccccc}\tilde{L} & = & A & L & B \\ {k\times k}&&k\times n&n\times n&n\times k\end{array} 
\end{equation}
where the matrices $A$ and $B$ are defined by
\begin{equation}
\label{lump_AB}
A_{li}=\begin{cases}\frac{\pi_i}{\sum_{i'\in\Gamma_l}\pi_{i'}},&i\in\Gamma_l,\\
0,&i\notin\Gamma_l,\end{cases}\quad\quad
B_{jr}=\begin{cases}1,& j\in \Gamma_r,\\
0,& j\notin\Gamma_r.\end{cases},\quad 1\le i,j\le n,~~1\le l,r\le k.
\end{equation}
In words,  $l$-th row of $A$ is the invariant distribution in $\Gamma_l$ and zeros elsewhere, while  $r$-th column of $B$ is the indicator function
of the connected component $\Gamma_r$.
One can readily check that $AB = I$. Furthermore, if $(\mu,x)$ is an eigenpair of $\tilde{L}$, then $(\mu,Bx)$ is an eigenpair of $B\tilde{L}A$, which
is a homogenized approximation to $L$.
Therefore, once we have computed an eigenpair $(\mu,x)$ of $\tilde{L}$, we can 
convert it to the corresponding eigenpair $(\mu,Bx)$ of the $n\times n$ matrix $B\tilde{L}A$ and use $Bx$ 
as the initial vector for the Rayleigh quotient iteration applied to the generator matrix $L$ of the full network.

The truncation procedure consists of two steps. First,  all edges $(i,j)$ with $V_{ij} > V_{\max}$ are removed from the network. 
Then the connected component of the network containing the global minimum $s^{\ast}_0$ is extracted.
If we need to combine truncation and lumping, we first truncate the network and then lump it. 
Note that the combination of truncation and lumping allows us, in principle, to obtain an approximation to any desired eigenpair.
Indeed, suppose we want to continue the asymptotic eigenpair $(A_k\exp(-\Delta_k/T),\phi^k_0)$ to a range of finite temperatures.
If we are unable to do it for the full network, we can pick
\begin{equation}
\label{vu}
 V_{\max} \ge \Delta_k + V_{s^{\ast}_k}\quad\text{ and}\quad \Delta_{\min}\le \Delta _k,
 \end{equation}
 truncate and lump the network,
and apply the Rayleigh Quotient iteration to the resulting symmetrized generator matrix with the appropriate initial approximation.
If $\Delta_{\min}  = \Delta _k$ and $V_{\max} = \Delta_k + V_{s^{\ast}_k}$, the resulting generator matrix will be $2\times 2$. 
Hence, its eigenvalue can be readily computed. However, it might be less accurate than we would like it to be. 
The farther
$\Delta_{\min}$ and $V_{\max}$ from their bounds,  the more challenging is the Rayleigh quotient iteration, but 
the better the computed eigenpair will approximate the exact one. 
Therefore, we pick $V_{\max}$ as large as possible and $\Delta_{\min}$ as small as possible
so that the Rayleigh quotient iteration still converges.


\subsubsection{The eigenvalue $\lambda(ICO-MARKS)$}
\label{sec:results}
\begin{figure}[htbp]
\begin{center}
\includegraphics[width = 0.6\textwidth]{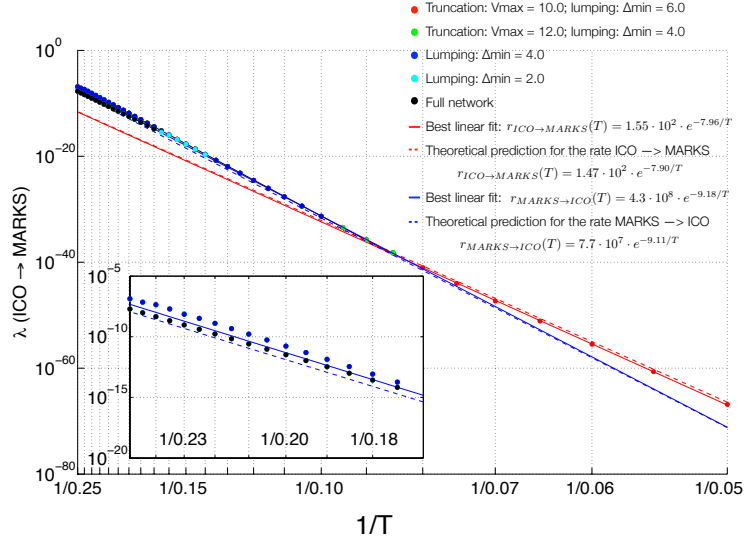}
\caption{The eigenvalue $ICO - MARKS$ corresponding to the transition process between 
the icosahedral and the Marks decahedral funnels in LJ$_{75}$.
}
\label{fig:con}
\end{center}
\end{figure}
In this Section, we discuss the behavior of the computed eigenvalue $\lambda(ICO-MARKS)(T)$  for $0.05\le T\le 0.25$ 
presented in Fig. \ref{fig:con}. We remind that the asymptotic estimate for this eigenvalue as $T\rightarrow 0$ is given by $1.472\cdot 10^2\exp(-7.897/T)$.
We were able to compute:
\begin{description}
\item[red dots:] an approximation to  $\lambda(ICO-MARKS)(T)$ using the truncated and lumped network 
with $V_{\max}=10.0$ and $\Delta_{\min} = 6.0$ for $0.05\le T\le 0.25$;
\item[green dots:] an approximation to  $\lambda(ICO-MARKS)(T)$ using the truncated and lumped network 
with $V_{\max}=12.0$ and $\Delta_{\min} = 4.0$ for $0.085\le T\le 0.25$;
\item[blue dots:] an approximation to  $\lambda(ICO-MARKS)(T)$ using the lumped network 
with $\Delta_{\min} = 4.0$ for $0.10\le T\le 0.25$;
\item[cyan dots:] an approximation to  $\lambda(ICO-MARKS)(T)$ using the lumped network 
with $\Delta_{\min} = 2.0$ for $0.14\le T\le 0.25$;
\item[black dots:] $\lambda(ICO-MARKS)(T)$ using the full network for $0.17\le T\le 0.25$.
The initial approximations for the eigenvectors  were obtained from the lumped network with $\Delta_{\min} = 2.0$
as described in Section \ref{sec:lump}.
\end{description}
The numbers of vertices $N$ and maximal vertex degrees $d_{\max}$ of the corresponding networks are listed in Table 2.
\begin{table}[htdp]
\caption{The number of vertices and maximal vertex degrees in the truncated and lumped network. $V_{\max}=\infty$ means no truncation.
$\Delta_{\min} = 0.0$ means no lumping. The last row corresponds to the full LJ$_{75}$ network.}
\begin{center}
\begin{tabular}{|c|c|c|c|}
\hline
$V_{\max}$&$\Delta_{\min}$&$N$ & $d_{\max}$\\
\hline
10.0&6.0 & 91 & 90\\
12.0&4.0 & 9361 & 6745\\
$\infty$ & 4.0 & 24833 & 13888\\
$\infty$& 2.0 & 55251 & 6663\\
$\infty$& 0.0 & 169523 & 740\\
\hline
\end{tabular}
\end{center}
\label{table2}
\end{table}%

The blue, green, and red dots in Fig. \ref{fig:con} visually coincide in the intersection of their  intervals $0.10\le T\le 0.25$, while 
the black and cyan dots visually coincide on the intersection of their intervals $0.17\le T\le 25$. The inset zooms in the plots
at the temperature range $0.17\le T\le 25$. The cyan and black dots are indistinguishable in it too. Beside that, the blue and the cyan dots 
are indistinguishable for $0.14\le T\le 0.16$.

The graph of the estimate of the eigenvalue $\lambda(ICO-MARKS)(T)$ versus $1/T$
in the logarithmic scale in the $y$-coordinate is nearly a piecewise linear curve consisting of two line segments (Fig. \ref{fig:con}). 
The best least squares fit with 
a linear function on the interval $0.05\le T\le 0.08$ (the solid red line)  
can be compared to the asymptotic estimate obtained by Algorithm 1 (the dashed red line):
\begin{align}
\text{Algorithm 1}:~~ \lambda_{4395}\equiv\lambda(ICO\rightarrow MARKS) &\approx 1.472\cdot 10^2e^{-7.897/T},\label{ico-marks}\\
\text{best linear fit}:~~ \lambda(ICO-MARKS) &\approx 1.552\cdot 10^2e^{-7.961/T}.\notag
\end{align}
For $T> 0.08$, the best least squares fit with a linear function (the solid blue line) (obtained for $0.08\le T\le 0.12$)
can be compared to the asymptotic estimate for the eigenvalue $
\lambda(MARKS\rightarrow ICO)$ { obtained pretending that the deepest minimum is located in the icosahedral funnel}: 
\begin{align}
\text{Prediction}:~~\lambda( MARKS\rightarrow ICO) &\approx 7.674\cdot 10^7e^{-9.107/T},\label{marks-ico}\\
\text{best linear fit}:~~\lambda(ICO-MARKS)  & \approx 4.307\cdot 10^8e^{-9.187/T}.\notag
\end{align}

{ Recall that
$T_{ss}^{LJ75}=0.08$ is the the solid-solid  transition} critical temperature.  
For $T<0.08$,
the system spends more time  in the Marks decahedral funnel rather than in the icosahedral one, while for $T>0.08$ it is the other way around.
This { critical} temperature is extremely low in comparison with $\Delta(ICO-MARKS)=7.897$.  As  Fig. \ref{fig:con} demonstrates, at $T<0.08$,
the asymptotic estimate for the eigenvalue is close to the estimate computed using truncated and lumped matrix with $V_{max} = 10.0$ and
$\Delta_{min} = 6.0$. For $T$ slightly larger than $T=0.08$, the system spends more time in the icosahedral funnel
with several local minima with close values of the potential at its bottom (minima  92, 93, 94, 95, 96, 97, 98, ...).
Effectively, they can be lumped together into one minimum with the value of the potential lower than that at minimum 1. This is why the 
predicted asymptotic estimate \eqref{marks-ico} 
 turns out to be quite accurate.
Due to the fact that the barrier $\Delta = 9.107$ separating  state 1 from the icosahedral funnel is very high in comparison with $T_{sl}^{LJ75}=0.25$,
this prediction remains quite accurate up to $T_{sl}^{LJ75}=0.25$ as shows the inset in Fig. \ref{fig:con}.


\subsubsection{The eigencurrent $F(ICO - MARKS)$}
\label{sec:ecur}
The computed eigenvector $\phi(ICO-MARKS)$ allows us to quantify the transition 
process between the Marks decahedral and icosahedral funnels 
in terms of distributions of the corresponding eigencurrent $F(ICO-MARKS)$.
Important characteristics of this process are the distributions of the emission and absorption of the eigencurrent, 
the location of the emission-absorption cut (the EA-cut), { and}
 the distribution of the eigencurrent in the EA-cut. Furthermore, if the eigencurrent happen to { focus} around 
 some subset of edges, one can extract this set of edges and hence identify major reactive channel or channels.

First{,} we establish the location of the emission-absorption cut as a function of temperature.
As the temperature increases, the location of the EA-cut along the MinMax path {(Fig. \ref{fig:lock})(a)}
shifts toward the icosahedral funnel.
At $T\rightarrow 0${,}  it must be at the edge $(25811,73992)$ corresponding to the highest potential barrier separating states 1 and 92.
Its height is 9.107 with respect to the potential at minimum 1. 
At $0.17\le T\le 0.235$ the emission-absorption cut intersects the MinMax path  at the edge $(3471,12596)$
corresponding to another barrier separating states 1 and 92 that is almost as high as the highest one: its height is 8.724 with respect to minimum 1.
At $T=0.24$, the emission-absorption cut further drifts toward minimum 92 to the edge $(20588,3472)$, 
and reaches the edge $(20587,20588)$ at $0.245\le T\le 0.25$.
This shift can be explained by the entropic effect: the icosahedral funnel is significantly wider than the Marks decahedral one. As the system  
overcomes the highest barrier at relatively high temperature ($T\ge 0.17$),  it is more likely for it to overcome it back than to find its way to the bottom 
of the icosahedral funnel due to its width and complexity. 
\begin{figure}[htbp]
\begin{center}
\centerline{
(a)\includegraphics[width=0.45\textwidth]{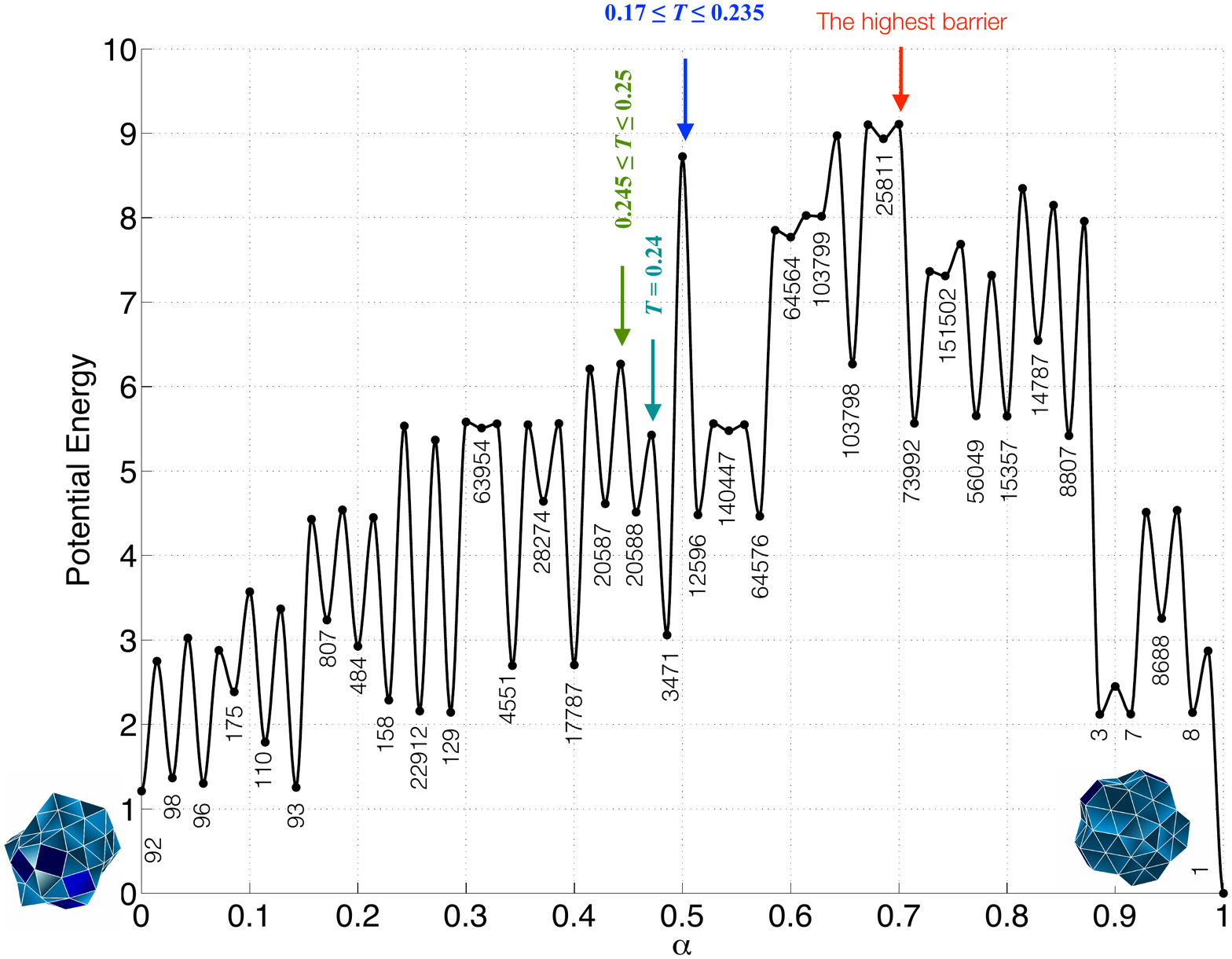}
(b)\includegraphics[width=0.45\textwidth]{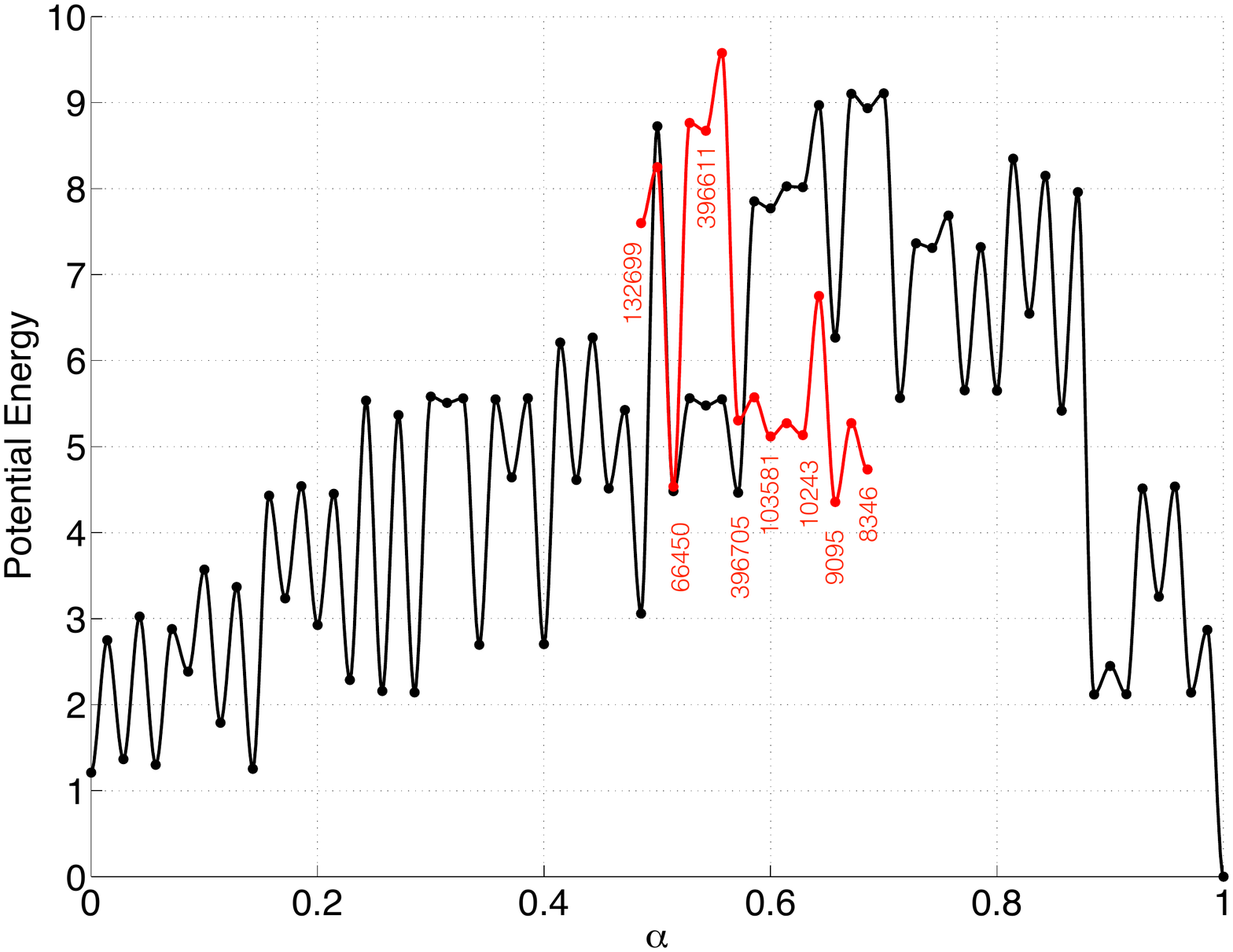}
}
\caption{
(a): The locations of the emission-absorption cut  for the eigencurrent $F(ICO-MARKS)$
along the MinMax path at temperatures $T\rightarrow 0$ and $0.17\le T\le 0.25$.
(b): The energy plot (red) along the sequence of edges along which the eigencurrent $F(ICO-MARKS)$
is focused at $0.17\le T\le 0.25$. 
}
\label{fig:lock}
\end{center}
\end{figure}
\begin{figure}[htbp]
\begin{center}
\includegraphics[width=0.6\textwidth]{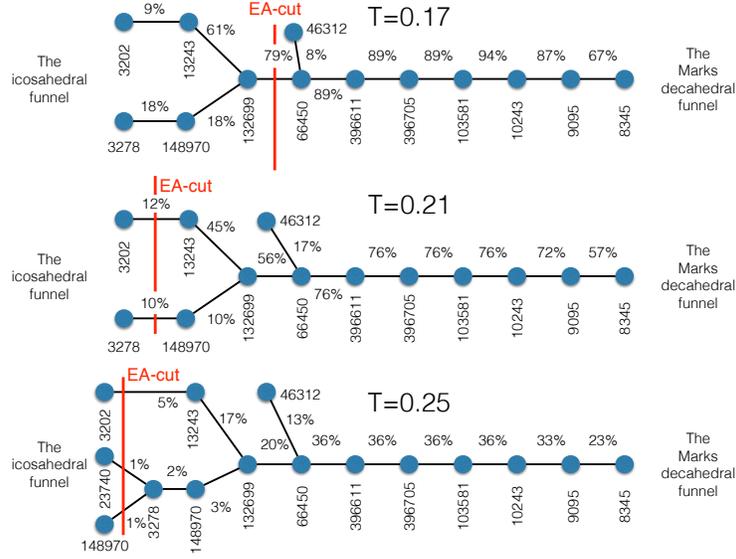}
\caption{
The network of edges passing through the
EA-cut along which the eigencurrent $F(ICO-MARKS)$ is concentrated.
}
\label{fig:chain}
\end{center}
\end{figure}

Comparing the values of the eigencurrent $F(ICO-MARKS)$  with its distribution in the EA-cut, 
we can extract a sequence of edges
$$
132699 \rightarrow 66450 \rightarrow 396611 \rightarrow 396705 \rightarrow 103581 \rightarrow 10243 \rightarrow 9095 \rightarrow 8345,
$$
 along which the eigencurrent $F(ICO-MARKS)$  is concentrated (see Fig. \ref{fig:chain}) at $0.17\le T\le 0.25$. 
 At $T=0.17$, about 90\% of the eigencurrent follows it. At $T=0.19$, this number reduces to 76\% and drops down to 36\% $T=0.25$.
This sequence is not a subsequence of the asymptotic zero-temperature path. The energy plot along it is shown in Fig. \ref{fig:lock}(b).
The EA-cut shifts toward the icosahedral funnel as the temperature increases. The eigencurrent widely spreads out in the icosahedral funnel.
So it does, but to a lesser extent, in the Marks decahedral funnel. 
The distributions of the eigencurrent $F(ICO-MARKS)$ in the EA-cut at $0.17\le T\le 0.25$ are shown in Fig. \ref{fig:ecur}(a). 
The eigencurrent is highly focused within the EA-cut  $0.17\le T\le 0.19$, while for $0.21\le T\le 0.25$, it is spread out. 
This is caused by the shift of the EA-cut toward the icosahedral funnel as well as the spreading of the eigencurrent due temperature increase.

The emission (solid curves) and absorption (dashed curves) distributions of the eigencurrent at $0.17\le T\le 0.25$
are shown in Fig. \ref{fig:ecur}(b). 
The intermediate asymptotics for these distributions in the log-log scale are straight lines, which means that they have heavy (power law) tails.
While nearly all  {eigencurrent} at $T=0.17$ is absorbed by two states: 1 and 239139, its emission is  spread out
across the icosahedral funnel. As the temperature increases, both distributions of emission and absorption widen (see Table 3). 
Fig. \ref{fig:ecur_q} visualizes the emission and absorption distributions in the plane of the bond-order parameters $(Q_4,Q_6)$.
The emitting and absorbing states can be separated, though not perfectly, in the $(Q_4,Q_6)$ - plane.
\begin{figure}[htbp]
\begin{center}
\centerline{
(a)
\includegraphics[width=0.45\textwidth]{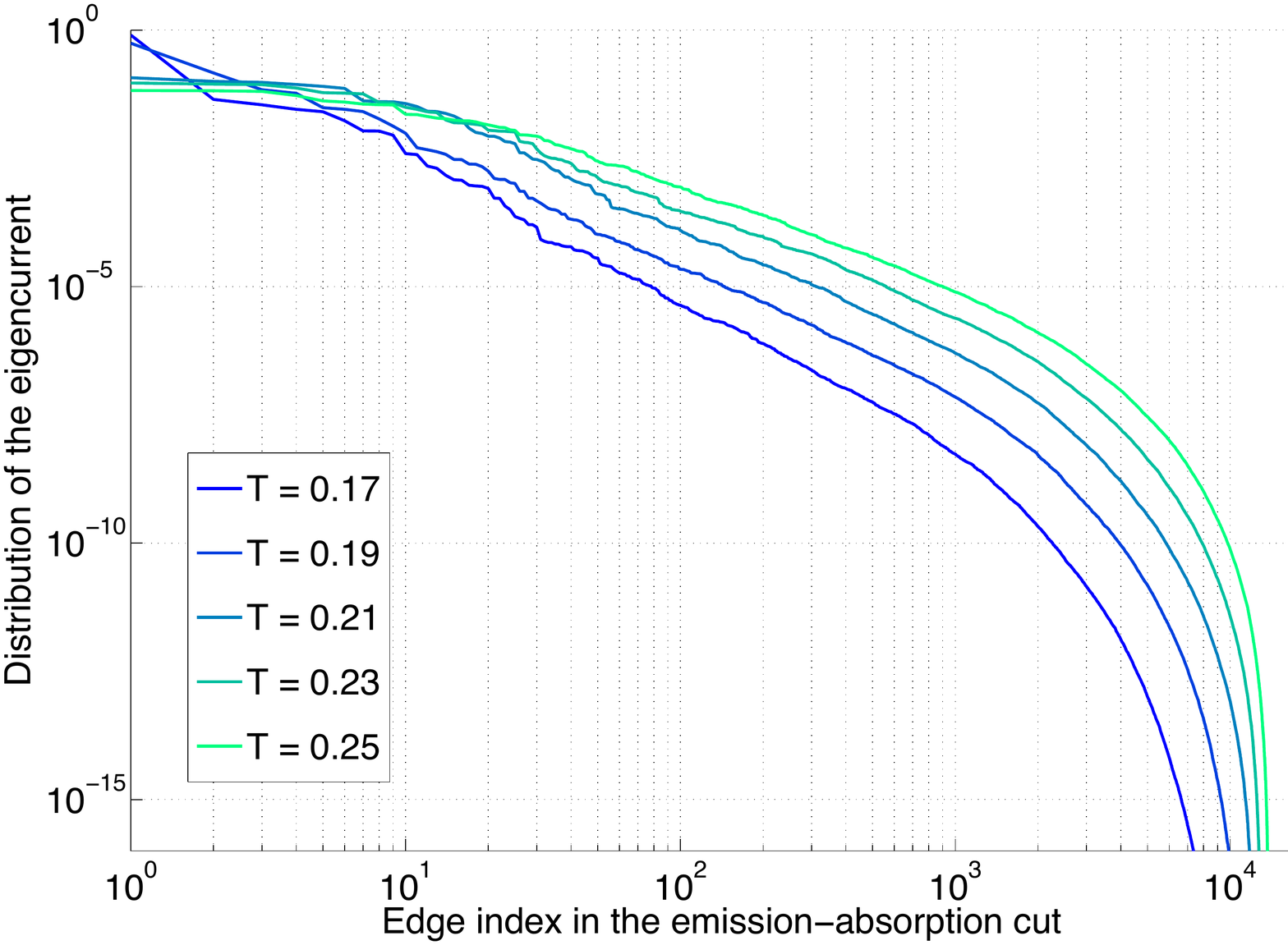}
(b)
\includegraphics[width=0.45\textwidth]{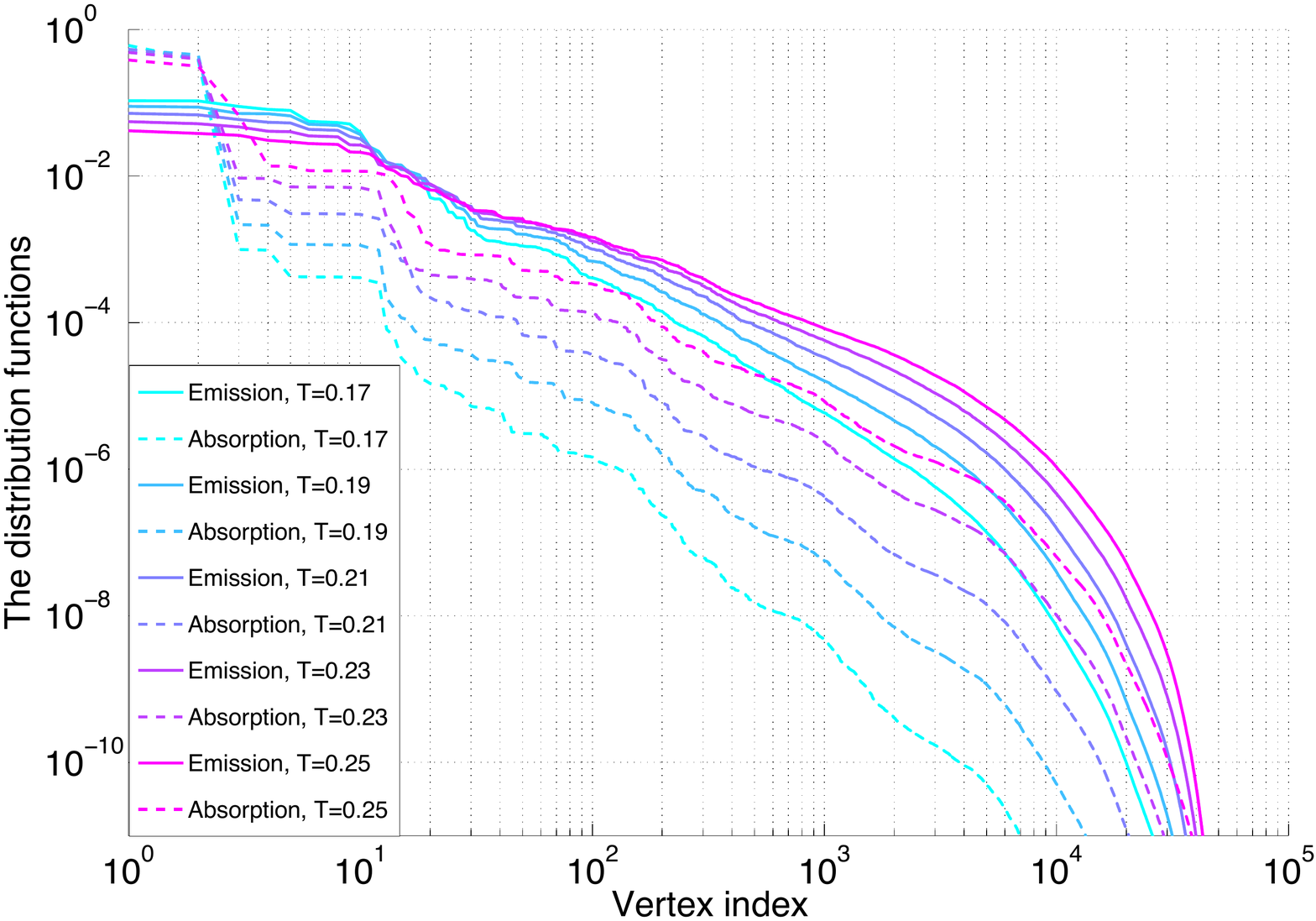}
}
\caption{
(a): The distributions of the eigencurrent $F(ICO-MARKS)$ in the emission-absorption cut at $0.17\le T\le 0.25$. (b): The  emission (solid curves) 
and absorption (dashed curves) distributions of the eigencurrent $F(ICO-MARKS)$
$0.17\le T\le 0.25$.
}
\label{fig:ecur}
\end{center}
\end{figure}

\begin{table}[htdp]
\caption{The numbers of states emitting ($N_e$) and absorbing ($N_a$) 90\% and 99\% 
of the eigencurrent $F(ICO-MARKS)$ }
\begin{center}
\begin{tabular}{|c|c|c|c|c|}
\hline
Temperature& $N_{e}(0.90)$ & $N_{e}(0.99)$ & $N_{a}(0.90)$ & $N_{a}(0.99)$\\
\hline
0.17 & 41 & 513 & 2 & 2\\
0.19 & 111 & 1437 & 2 & 6 \\
0.21 & 270 & 2941 & 2 & 20\\
0.23 & 646 & 4741 & 4 & 145\\
0.25 & 1342 & 6737 & 16 & 862\\
\hline
\end{tabular}
\end{center}
\label{table3}
\end{table}%

\begin{figure}[htbp]
\begin{center}
\centerline{
(a)
\includegraphics[width=0.45\textwidth]{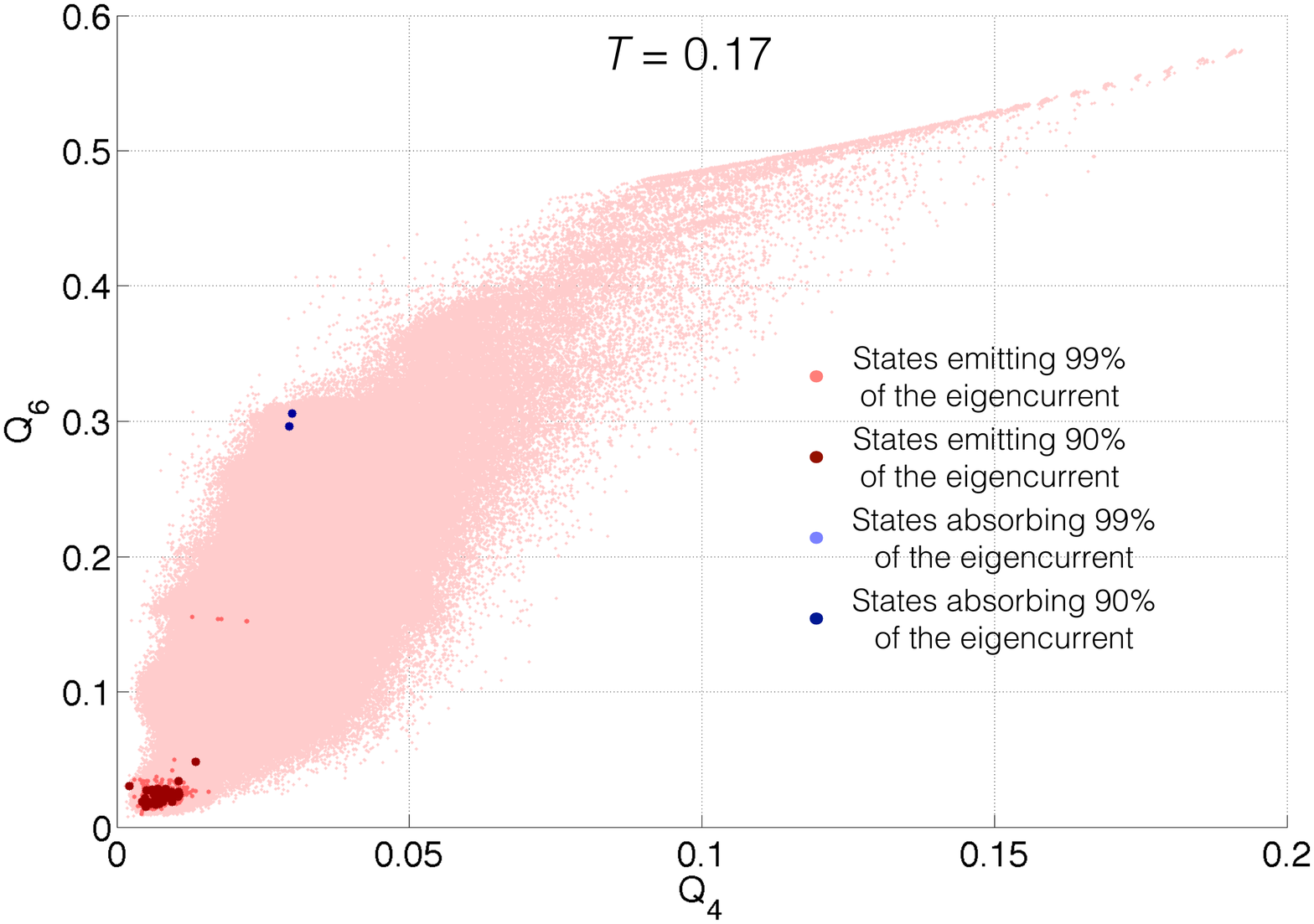}
(b)
\includegraphics[width=0.45\textwidth]{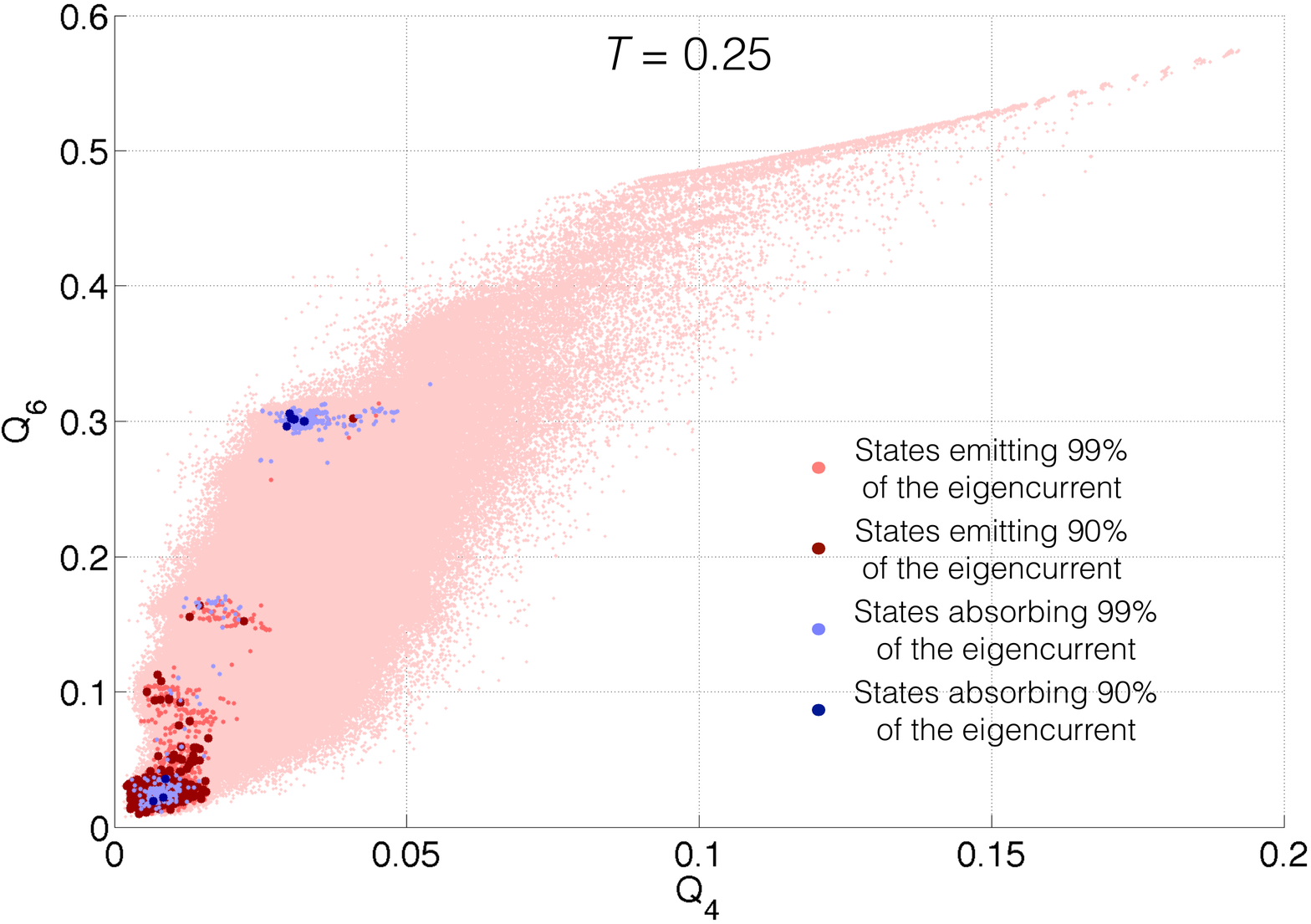}
}
\caption{
(a): States emitting and absorbing most of the eigencurrent  $F(ICO-MARKS)$ at $T=0.17$ in the bond-order parameters $Q_4$ and $Q_6$.
(b): States emitting and absorbing most the eigencurrent  $F(ICO-MARKS)$ at $T=0.25$ in the bond-order parameters $Q_4$ and $Q_6${.}
}
\label{fig:ecur_q}
\end{center}
\end{figure}


\section{Concluding remarks}
\label{sec:conclusion}
We have presented the two - step approach for spectral analysis of large stochastic networks.
Step one is the asymptotic analysis and step two is the finite temperature continuation.
Its power is demonstrated on the large and complex stochastic network representing  the energy landscape of LJ$_{75}$.

The theoretical justification for the asymptotic part of this approach 
is completed for time-reversible networks. 
For time-irreversible networks, we have extended Wentzell's result \cite{wentzell2,f-w} obtained 
sharp estimates for the eigenvalues (Eqs. \eqref{eq3}-\eqref{eq5}), i.e., containing both exponents and pre-factors. 
Algorithm 1 is equally  efficient for both time-reversible and time-irreversible 
continuous-time Markov chains. We have proven the nested properties of the optimal W-graphs (Theorem \ref{the:nested})
justifying the construction implemented in Algorithm 1.
We are planning to address the problem of obtaining asymptotic estimates for left and right eigenvectors in the future.

The finite temperature continuation technique used for LJ$_{75}$ 
is quite general and can be used for an arbitrary time-reversible network of the considered type.
 The combination of truncation and lumping can fight floating point arithmetic issues and enable us 
 to obtain approximations to eigenvalues and eigenvectors. These approximations for eigenvectors,  in turn,
 can be used for obtaining initial guesses for eigenvectors for the Rayleigh quotient iteration applied to 
a less lumped and less truncated network. 
And so on, until we are able to compute eigenvalues and eigenvectors for the full network.
 
 The  continuation technique is yet to be extended to the time-irreversible case
 where the generator matrix is not symmetrizable and eigenpairs might become complex .
 
 Our analysis of LJ$_{75}$ can be compared to the one of LJ$_{38}$ \cite{cam1,cve,cspec1,cspec2}.
 Both of these networks represent energy landscapes with two major funnels one of which is deep and narrow and the other one is wide
 and not as deep. Both of these networks are large and complex  and there is no scale separation between various transition
 processes in them. This causes the observed absence of spectral gaps (compare Fig. \ref{fig:LJ75deltas} to Fig. 6 in \cite{cspec1}).
 For both of these networks, the transition process{es} between two major funnels become diverse as the temperature increases, and the distributions
 of the corresponding eigencurrents in the corresponding EA-cut look similar (compare Fig. \ref{fig:ecur}(a) to Fig. 8 in \cite{cspec2}).
 For both of these distributions{,} the intermediate asymptotics are power laws.
 
 However, there is an important difference in the qualitative behavior of the 
 eigenvalues corresponding to the transition process between the two major funnels of LJ$_{75}$ and LJ$_{38}$.
 {It is caused by} the fact that the solid-solid phase transition temperature  in LJ$_{75}$ is lower than that in LJ$_{38}${,}  
 while the potential barrier separating the two funnels in LJ$_{75}$ is approximately twice as high as that in LJ$_{38}$.
In LJ$_{75}$, the graph of the eigenvalue $\lambda(ICO-MARKS)$ versus $1/T$ in the logarithmic scale
 is a piecewise linear curve { with a kink} at $T_{ss}^{LJ75}$ (Fig. \ref{fig:con})
 { and linear pieces nearly perfectly agreeing with the theoretical estimates.} 
In LJ$_{38}$, the graph of the eigenvalue $\lambda(ICO-FCC)$ versus $1/T$ in the logarithmic scale is a smooth curve, 
because the transition process dramatically broadens before $T_{ss}^{LJ38}$ is reached (Fig. 7 in \cite{cspec2}).


\section{Acknowledgements}
\label{sec:ac}
We thank Professor David Wales for providing us with the data for the LJ$_{75}$ network and valuable discussion.
We are grateful to Mr. Weilin Li for valuable discussions at the early stages of the development of Algorithm 1.
This work was partially supported by NSF grant 1217118.

  \renewcommand{\theequation}{A-\arabic{equation}}
  \setcounter{equation}{0}  
  \section*{APPENDIX}  
  {\bf Proof of Theorem \ref{the:nested}.}
For brevity, we will denote the set of arcs with tails belonging to a subset of states $\tilde{S}$ in the W-graph $g$ by $\mathcal{A}(g;\tilde{S})${,} and the sum of weights
of these arcs by $\Sigma(g;\tilde S)$, i.e.,
$$
\mathcal{A}(g;\tilde{S}): = \{(i\rightarrow j)\in g~|~ i\in \tilde{S}\},\quad \Sigma(g;\tilde{S}):=\sum_{\mathcal{A}(g;\tilde{S})}U_{ij}.
$$

\begin{proof}
First we prove Claims 1 and 2.
 Since there are $k$ sinks in $g^{\ast}_{k}$ and $k+1$ connected components in $g^{\ast}_{k+1}$, at least one connected component of $g^{\ast}_{k+1}$
contains no sink of $g^{\ast}_k$. 
Pick {one} such  connected component of $g^{\ast}_{k+1}$ and denote { its set of vertices and its sink by $S_k$ and  $s^{\ast}_k$ respectively}. 
\item The subset of vertices $S\backslash S_k$ contains $k$ sinks of $g^{\ast}_{k+1}$. Hence 
$$
|\mathcal{A}(g^{\ast}_{k+1};S\backslash S_k)| = |S\backslash S_k| - k.
$$
Since $S_k$ contains no sinks of $g^{\ast}_k$, all $k$ sinks of $g^{\ast}_k$ lie in $S\backslash S_k$. Hence  
$$
|\mathcal{A}(g^{\ast}_{k};S\backslash S_k)| = |S\backslash S_k| - k = |\mathcal{A}(g^{\ast}_{k+1};S\backslash S_k)|.
$$
Since the W-graphs $g^{\ast}_{k+1}$ and $g^{\ast}_k$ are optimal, we have that
$$
\Sigma(g^{\ast}_{k+1};S\backslash S_{k}) = \Sigma(g^{\ast}_{k};S\backslash S_{k})  = \min_{g\in G(k;S\backslash S_k)}\sum_{(i\rightarrow j)\in g} U_{ij},
$$
where $G(k;S\backslash S_k)$ denotes the set of all W-graphs with $k$ sinks defined on the set of vertices $S\backslash S_k$.
Since the optimal W-graphs $g^{\ast}_{k+1}$ and $g^{\ast}_k$ are unique, we conclude that
$$
\mathcal{A}(g^{\ast}_{k};S\backslash S_k) = \mathcal{A}(g^{\ast}_{k+1};S\backslash S_k).
$$
This implies that the sinks in $S\backslash S_k$ of $g^{\ast}_k$ and $g^{\ast}_{k+1}$ coincide.
Hence, the connected component {$S_k$} of $g^{\ast}_{k+1}$ containing no sinks of $g^{\ast}_k$ 
is unique and every sink of $g^{\ast}_k$ is also a sink of $g^{\ast}_{k+1}$.
Thus, Claims 1 and 2 are proven.

Now we prove Claim 3. 
Suppose that there are $l>1$ arcs with tails in $S_k$ and heads in $S\backslash S_k$ in $g^{\ast}_{k}$. Then the set of vertices $S_k$
can be subdivided into two subsets, $\tilde{S_k}$ and $S_k\backslash \tilde{S}_k$, so that
$\tilde{S}_k$  is connected in $g^{\ast}_k$, contains the sink $s^{\ast}_k$,
 and $S_k\backslash \tilde{S}_k$ and $\tilde{S}_k$ are not connected in $g^{\ast}_k$, as shown in Fig. \ref{fig:claim3}.
 \begin{figure}[htbp]
\begin{center}
\includegraphics[width=0.6\textwidth]{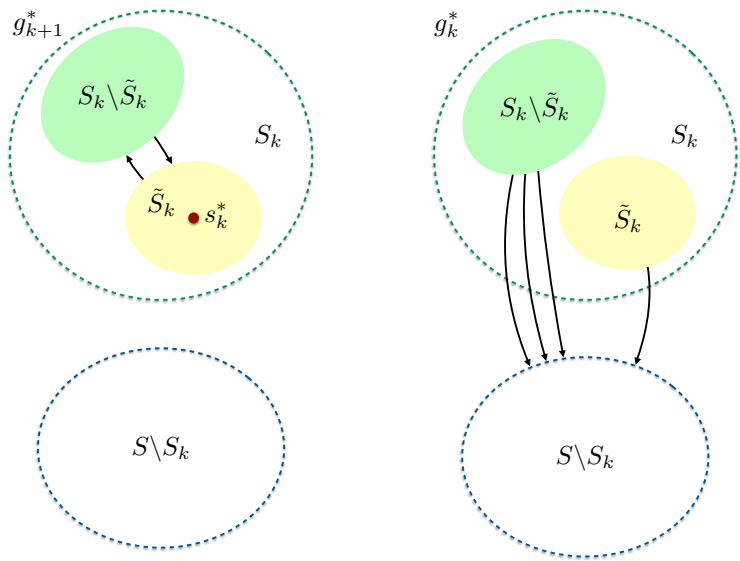}
\caption{An illustration for the proof of Claim 3 of Theorem \ref{the:nested}.}
\label{fig:claim3}
\end{center}
\end{figure}

By construction, $S_k\backslash\tilde{S}_k$ contains no sinks of $g^{\ast}_{k+1}$, hence 
$$
|\mathcal{A}(g^{\ast}_{k+1};S_k\backslash\tilde{S}_k)| = |S_k\backslash\tilde{S}_k|,
$$
 i.e., 
every vertex in $S_k\backslash\tilde{S}_k$ has an outgoing arc.
Since all directed paths in $g^{\ast}_{k+1}$ starting in the vertices of $S_k\backslash\tilde{S}_k$ lead to the sink $s^{\ast}_k$, 
$$
\mathcal{A}(g^{\ast}_{k+1};S_k\backslash\tilde{S}_k)\neq \mathcal{A}(g^{\ast}_{k};S_k\backslash\tilde{S}_k).
$$
 However, the set $S_k\backslash\tilde{S}_k\subset S_k$ contains no sinks of $g^{\ast}_k$. Hence
 $$
|\mathcal{A}(g^{\ast}_{k};S_k\backslash\tilde{S}_k)| = |S_k\backslash\tilde{S}_k| = |\mathcal{A}(g^{\ast}_{k+1};S_k\backslash\tilde{S}_k)|.
$$
Suppose we replace $\mathcal{A}(g^{\ast}_{k+1};S_k\backslash\tilde{S}_k)$ with $\mathcal{A}(g^{\ast}_{k};S_k\backslash\tilde{S}_k)$
in $g^{\ast}_{k+1}$. 
We create no cycles by this replacement, because $(i)$
there is no arc from $S_k\backslash\tilde{S}_k$ to $\tilde{S}_k$ in $g^{\ast}_k$,
and $(ii)$
there is no arc from $S\backslash S_k$ to $S_k$ in $g^{\ast}_{k+1}$.
Since $g^{\ast}_{k+1}$ is the unique  optimal W-graph with $k+1$ sinks, we have:
\begin{equation}
\label{A1}
\Sigma(g^{\ast}_{k+1};S_k\backslash\tilde{S}_k) < \Sigma(g^{\ast}_{k};S_k\backslash\tilde{S}_k).
\end{equation}
On the other hand, suppose we replace   $\mathcal{A}(g^{\ast}_{k};S_k\backslash\tilde{S}_k)$ with  
$\mathcal{A}(g^{\ast}_{k+1};S_k\backslash\tilde{S}_k)$ in $g^{\ast}_k$. Since there is no arc from $\tilde{S}_k$ to $S_k\backslash\tilde{S}_k$
in $g^{\ast}_k$, this replacement creates no cycle. 
Since $g^{\ast}_{k}$ is the unique  optimal W-graph with $k$ sinks, we have:
\begin{equation}
\label{A2}
\Sigma(g^{\ast}_{k};S_k\backslash\tilde{S}_k) > \Sigma(g^{\ast}_{k+1};S_k\backslash\tilde{S}_k).
\end{equation}
Eq. \eqref{A1} contradicts to Eq. \eqref{A2}. Hence the assumption that there is more than one outgoing arc from $S_k$ to $S\backslash S_k$ in $g^{\ast}_k$
leads to a contradiction. This proves Claim 3.

\end{proof}

\end{document}